\documentclass[utf8]{frontiersSCNS} 

\usepackage{url,hyperref,microtype,subcaption}
\usepackage[onehalfspacing]{setspace}
\usepackage{amsmath,amssymb}
\usepackage{comment}

\newcommand{\nuc}[2]{\ensuremath{{}^{#1}\mathrm{#2}}}
\newcommand{\C}[1]{\nuc{#1}{C}}
\newcommand{\He}[1]{\nuc{#1}{He}}
\newcommand{\p}{\ensuremath{\mathrm{p}}}

\newcommand{\reactionBan}{\ensuremath{\nuc{11}{B}(\alpha,\mathrm{n})\nuc{14}{N}}}
\newcommand{\reactionCan}{\ensuremath{\C{13}(\alpha,\mathrm{n})\nuc{16}{O}}}
\newcommand{\reactionCpg}{\ensuremath{\C{13}(\p,\gamma)\nuc{14}{N}}}
\newcommand{\reactionCag}{\ensuremath{\C{12}(\alpha,\gamma)\nuc{16}{O}}}
\newcommand{\reactionCC}{\ensuremath{\C{12}+\C{12}}}
\newcommand{\reactionCCpNa}{\ensuremath{\C{12}(\C{12},\p{})\nuc{23}{Na}}}
\newcommand{\reactionCCaNe}{\ensuremath{\C{12}(\C{12},\alpha)\nuc{20}{Ne}}}
\newcommand{\reactionNpg}{\ensuremath{\nuc{14}{N}(\p,\gamma)\nuc{15}{O}}}
\newcommand{\reactionNean}{\ensuremath{\nuc{22}{Ne}(\alpha,\mathrm{n})\nuc{25}{Mg}}}
\newcommand{\reactionNeag}{\ensuremath{\nuc{22}{Ne}(\alpha,\gamma)\nuc{26}{Mg}}}

\newcommand{\fm}{\ensuremath{\,\mathrm{fm}}}

\newcommand{\mueV}{\ensuremath{\,\mathrm{\mu{}eV}}}

\newcommand{\kV}{\ensuremath{\,\mathrm{kV}}}
\newcommand{\keV}{\ensuremath{\,\mathrm{keV}}}

\newcommand{\MV}{\ensuremath{\,\mathrm{MV}}}
\newcommand{\MeV}{\ensuremath{\,\mathrm{MeV}}}
\newcommand{\Coulomb}{\ensuremath{\,\mathrm{C}}}
\newcommand{\degC}{\ensuremath{{}^\circ\mathrm{C}}}

\newcommand{\etal}{\emph{et al.}}
\newcommand{\ie}{\emph{i.\,e.}}

\newcommand{\range}[2]{\ensuremath{#1 - #2}}
\newcommand{\LUNAfourhundred}{LUNA\,400}

\def\keyFont{\fontsize{8}{11}\helveticabold }
\def\firstAuthorLast{Sample {et~al.}} 
\def\Authors{Ferraro F.\,$^{1}$, Ciani G.F.\,$^{2,3}$, Boeltzig A.\,$^{3}$, Cavanna F.\,$^{4}$ and Zavatarelli S.\,$^{1,*}$}

\def\Address{$^{1}$ Università degli Studi di Genova and INFN, Sezione di Genova, Via Dodecaneso 33, 16146 Genova, Italy \\
$^{2}$ Dipartimento di Matematica e Fisica, Università della Campania L. Vanvitelli, Viale Lincoln 5, Caserta,  Italy \\
$^{3}$ INFN, Laboratori Nazionali del Gran Sasso (LNGS), Via G. Acitelli 22, 67100 Assergi, Italy \\
$^{4}$ INFN, Sezione di Torino, Via P. Giuria 1, 10125 Torino, Italy }

\def\corrAuthor{Corresponding Author}

\def\corrEmail{sandra.zavatarelli@ge.infn.it}

\begin{document}
\onecolumn
\firstpage{1}

\title[Running Title]{The study of key reactions shaping the post main-sequence evolution of massive stars in underground facilities} 

\author[\firstAuthorLast ]{\Authors} 
\address{} 
\correspondance{} 

\extraAuth{}

\maketitle

\begin{abstract}

The chemical evolution of the Universe and several phases of the stellar life are regulated by minute nuclear reactions. The key point for each of these reactions is the value of cross sections at the energies at which they take place in stellar environments.
Direct cross-section measurements are mainly hampered by the very low counting rate and by cosmic background, nevertheless they have become possible by combining the best experimental techniques with the cosmic silence of an underground laboratory.
In the nineties the LUNA (Laboratory for Underground Nuclear Astrophysics) collaboration opened the era of underground nuclear astrophysics installing first a home-made $50\kV{}$ and later on, a second $400\kV{}$ accelerator under the Gran Sasso mountain in Italy: in 25 years of experimental activity, important reactions responsible for hydrogen burning could have been studied down to the relevant energies thanks to the high current proton and helium beams provided by the machines.
The interest to the next and warmer stages of star evolution (\ie{} post main sequence, helium and carbon burning) drove a new project based on an ion accelerator in the MV range called LUNA-MV, able to deliver proton, helium and carbon beams. The present contribution is aimed to discuss the \textcolor{red}{state-of-the-art} for some  selected key processes of post main sequence stellar phases: the \reactionCag{} and the \reactionCC{} fundamental for helium and carbon burning phases, and  \reactionCan{} and \reactionNean{} that are relevant to the synthesis of heavy elements in AGB stars. The perspectives opened by an underground MV-facility will be highlighted.

\tiny
 \keyFont{ \section{Keywords:} Underground nuclear astrophysics, helium burning, carbon burning, gamma spectroscopy, neutron spectroscopy} 
\end{abstract}

\section{Introduction}

The hypothesis that the energy which powers the Sun comes from thermonuclear reactions seems to be mainly due to  \cite{Eddington:1920} and Aston. After the discovery of nuclear reactions by Rutherford in the twenties it became clear that only the enormous amount of energy stored in the nuclei and released during fusion reactions was able to support the sun luminosity for a time period compatible with the geological datings (\cite{Weizsacker:1938},  \cite{Bethe:1938}): by fact, in order to properly understand the chemical evolution and the stellar energy engine, it is fundamental to precisely know how light nuclei are converted to heavier ones.

According to current theories, the first nuclei were formed through a network of nuclear reactions in the Big Bang nucleosynthesis (BBN), a few minutes after the Big Bang. BBN left our universe containing about 75\% hydrogen, 24\% helium by mass, with some traces of lithium and deuterium. The composition of the present Universe is not very different from the primordial one, \textcolor{red}{with}
the total mass elements heavier than hydrogen and helium ("metals" according to the astronomers) at the level of a few percent. Stars fuse light elements to heavier ones in their cores,  up to iron and nickel in the more massive stars. 

The most important stellar properties that determine the evolutionary fate of a star are its mass and its composition (\cite{Rolfs1988}, \cite{Iliadis-Wiley-2015}) : the larger the mass, the larger the temperature in the core. The star composition influences which reactions dominate the burning processes.

When a low-mass star like the Sun runs out of hydrogen in the core, it becomes a red giant star, fusing H to He via the CNO cycle in a shell surrounding an inert He core. When the core temperature reaches 100 million K, the He nuclei  in the core have sufficient kinetic energy to fuse to C (helium burning), forming \C{12} in a two-stage process. Subsequent fusion with another helium nucleus produces \nuc{16}{O} nuclei. This process, in symbols \reactionCag{}, is the main source of the carbon and oxygen found in the Universe, including that in our bodies and represents by fact the "Holy Grail" of nuclear astrophysics
since  the C/O ratio at the end of helium burning greatly affects the subsequent evolution of the star.
At some point, when the He in the core is exhausted, the stars start to burn He in a shell surrounding the inert C/O core, in addition to burning H to He in a shell surrounding the He burning region. This phase, referred to as the asymptotic giant branch (AGB), is characterised by thermal instabilities: at a given time the burning shells extinguish and the low-mass star will end its existence as a white dwarf, consisting mainly of C and O and supported by electron degeneracy pressure. 

Massive stars evolve very differently from low-mass stars. After the end of a burning phase, the  core contracts gravitationally, and the temperature increase can be sufficient to ignite the next and heavier nuclear fuel.
In case of masses larger that  11 \(M_\odot\), after undergoing He burning, the core experiences further burning episodes referred to as C-, Ne-, O- and Si-burning. The duration of each subsequent nuclear burning phase decreases significantly. There are two main reasons:  the first is that each burning phase releases by far less energy per unit mass with respect to the previous phase; the second that an increasing fraction of energy is radiated away by neutrinos. Therefore, while H burning may continue for many million years, C burning typically lasts hundreds of days and Si burning may run out in just one day.
After the last advanced burning stage (Si burning) the core consists mainly of iron isotopes: no more energy can be generated through fusion reactions. The core contracts and when it exceedes the Chandrasekhar mass limit, it collapses until the density of nuclear matter is reached. As a consequence of the neutron degeneracy pressure, the core rebounds and produces an outgoing shock wave. The wave heats and compresses the overlying layers of the star, consisting of successive shells of Si, O, Ne and C thus more episodes of nucleosynthesis, referred to as explosive Si-, O-, Ne- and C-burning, take place.

The creation of elements heavier than iron occurs mainly through neutron capture processes, eventually followed by beta decays in the so called  \textit{s}~(slow)-process (\cite{Kappeler-2011-RMP}) and \textit{r}~(rapid)-process. The \textit{r}-process dominates in environments with higher free neutrons fluxes and it produces heavier elements and more neutron-rich isotopes than the \textit{s}-process. Supernovae explosions and neutron star mergers are potential sites for the \textit{r}- process.
The \textit{s}-process  is slow in the sense that there is enough time for beta decays to occur before another neutron is captured: a network of reactions produces stable isotopes by moving along the valley of beta-decay stable isobars. This process primarily occurs within ordinary stars, particularly AGB stars, where the neutron flux is sufficient to cause neutron captures to recur every 10–100 years, much slower than for the \textit{r}-process, which requires 100 captures per second.

 The key point for each of these reactions is the value of cross sections at the energies at which they take place in stellar environments.
For most stellar scenarios, the  changes in the system are slower than the collision time between the ions or atoms inside the stars, thus the temperature profile is well-defined: the thermonuclear reaction rate depends on the Maxwell-Boltzmann velocity distribution and on the cross section $\sigma$(E) energy dependence (\cite{Rolfs1988}). Typical stellar temperatures for main-sequence low-mass stars, correspond to peak energies of the Maxwell-Boltzmann distribution of $k_B\,T \sim 0.9-90\keV{}$. In case of more massive stars during advanced burning stages, peak energies can be as high as few MeV. 
For charged particles induced reactions these energies are typically well below the Coulomb barrier due to the nuclei electrostatic repulsion and the nuclear reactions proceed via tunnel effect. As a consequence, the low values of the cross sections, ranging from pico- to femto-barn and even below, prevent their measurements in a laboratory at the Earth's surface where the signal to background ratio is too small mainly because of cosmic rays. The observed energy dependence of the cross section at high energies is extrapolated to astrophysically relevant energies leading to substantial uncertainties. In particular, the reaction mechanism might change, or there might be the contribution of unknown resonances which could completely dominate the reaction rate at the stellar energies.

In the nineties the LUNA collaboration proved that the installation of the experiments in a deep underground laboratory, the Gran Sasso National Laboratory, is a successful approach:  for the first time nuclear astrophysics measurements with very small counting rates, down to few events per month became a reality.

The  high current hydrogen and helium beams provided by the 50 kV (\cite{GREIFE1994327}) and, later on, by the LUNA-400 kV accelerators (\cite{Formicola03-NIMA})  allowed to investigate for the first time at stellar energies the most important reactions responsible for the hydrogen burning in the Sun, such as the  
$\nuc{3}{He}(\nuc{3}{He},2p)\nuc{4}{He}$
 (\cite{Bonetti:1999yt}) and for the BBN such as the 
 $\nuc{2}{H}(\p,\gamma)\nuc{3}{He}$
  (\cite{Casella02-NPA,Mossa:2020qgj}).

Full descriptions of LUNA and of the several results obtained in 25 years of experimental activity can be found in recent review papers (\cite{Broggini18-PPNP,Cavanna18-IJMPA,BrogginiWP2019}).

Such achievements have motivated two proposals for similar facilities in China (\cite{juna_collaboration_progress_2016}) and in the United States (\cite{CASPAR}).

The  importance to extend such precise studies to the processes relevant to the late
 and warmer stages of star evolution (post main-sequence phases,  helium and carbon burning)  has motivated the LUNA collaboration to acquire a new and more powerful  3.5\,MV single-ended accelerator . The new machine will deliver ion beams  of H$^+$, \He{4}$^+$, \C{12}$^+$ and \C{12}$^{++}$ in the energy range from 0.350 to 7\MeV{} with $100\,\mathrm{\mu{}A}$ - $1\,\mathrm{mA}$ intensity currents, depending on the ion species and on the energy value.

In the following sections, first we will  focus on the technical aspects which are important for an underground nuclear astrophysics experiment. 
Then,  the state of the art and the expected improvements from underground measurements are presented for some selected key processes of post-main-sequence stellar phases:  in detail, the \reactionCan{} and \reactionNean{}, that are sources of neutrons for the s-process in asymptotic giant branch stars (AGB) and during hydrostatic evolution of massive stars and the \reactionCag{} and the \reactionCC{}  reactions, key processes of helium and carbon burning, respectively.

In the conclusions,  the commissioning phase of the new accelerator  will be detailed, together with highlights about the exciting perspectives opened by the new facility in a larger time window scenario.

\section{The MV facility at Gran Sasso}
The MV facility will be hosted in the north side of Hall B in the Gran Sasso Laboratory  and will consist of an accelerator room with concrete walls and a multistory building housing the control room and technical facilities. The concrete walls and ceiling (thickness of 80 cm) of the accelerator room will act as neutron shielding.

 Nuclear astrophysics experiments require both high beam currents and a well-defined and stable beam energy: to perform reliable energy scans of the targets the accelerator terminal voltage must be stable to $<1\keV{}$ over many hours and to $<0.1\keV{}$ over one hour. A precise energy value is mandatory because of the almost exponential energy dependence of the cross section induced by the tunnel effect probability: a small fluctuation of the beam energy would cause a large uncertainty in the measured cross section value. Since for some reaction long data taking times are expected, the ion source must be able to run stably overnight without human intervention.

A 3.5\MV{} linear DC accelerator was specifically developed by  High Voltage Engineering to meet the stringent requirements on beam intensity and stability (\cite{Sen2019}).
The machine will deliver ion beams into two different beam lines  via a 35$^\circ$ switching analyzing magnet. Two independent target stations for solid and gas targets will be located at 2\,m distance from the analyzing magnet. The LUNA-MV accelerator is single-ended, \ie{} it has an ion source and an injector block located inside the accelerator tank in the high-voltage terminal. 

The need for high-intensity protons as well as carbon ions in the 2$^{+}$ charge state  were the reasons to prefer an  electron cyclotron resonance  (ECR) ion source for the accelerator. 

The accelerator operates at a terminal voltage (TV) range of \range{300\kV{}}{3.5\MV{}}, while the ion source can operate at \range{30\kV{}}{40\kV{}}. In the present system, high-intensity beam currents should be maintained over a large dynamic range: by considering a 1 mA current capability in case of a proton beam, the beam power can be as high as 3.5 kW. 
To guarantee voltage stability for longer time periods ($>$1 h) a high precision, low- temperature coefficient ($<$ 5 ppm \degC{}) resistor chain is used to measure the terminal voltage. 
Beam intensity on target for H, He and C ions are reported in Table 1. Compared to previous Singletron accelerators, the LUNA-MV has improved specifications for terminal voltage stability and ripple ($10^{-5}$). Beam energy reproducibility is in the order of $10^{-4}$.
A detailed description can be found in \cite{Sen2019}.

\begin{table*}[h]
\caption{Beam intensity on target}
\begin{center}
\begin{tabular}{|c|c|}
\hline
Ion Species & Current (\textit{e}$\mu$A)  \\

$ $ & \textit{TV~range}: \range{3.5}{0.5}\MV{} (\range{0.5}{0.3}\MV{}) \\
\hline
\nuc{1}{H}$^{+}$	& 1000 (500)	 \\
\nuc{4}{He}$^{+}$	& 500 (300)	 \\
\C{12}$^{+}$	& 150 (100)	 \\
\C{12}$^{++}$	& 100 (60)	 \\
\hline
\end{tabular}
\end{center}
\label{table_1}
\end{table*}

For practical considerations, targets for direct measurements of nuclear cross sections on stable nuclides are typically either in solid or gaseous state. The basic aspects of such targets are similar for experiments underground and on surface, but certain requirements are emphasized for experiments deep underground to fully embrace the advantages of the location.
In the case of a solid target, the beam energy loss occurs in a relatively small volume. The resulting power density, up to on the order of \range{10^2}{10^3}\,W/cm$^2$ at \LUNAfourhundred{} if the beam is stopped in the target, requires the target to be cooled to avoid an increase of temperature that would damage the target or accelerate beam-induced target degradation. For targets on an inert backing material, such as those produced by evaporation, sputtering or implantation, water cooling behind the target is often used to dissipate the heat. The maximal power densities attainable on target will increase with the next generation of underground accelerators, either because of higher beam energies at comparable intensities (such as the MV facility at Gran Sasso), or due to further increased beam intensities (cf.\ JUNA \cite{juna_collaboration_progress_2016}) compared to \LUNAfourhundred{}. Efforts are underway to adopt and advance techniques from surface experiments, such as cooling for high-powered targets (\cite{wolfgang_hammer_scorpion_1986}) or large-area reaction targets (\cite{chen_preparation_2020}), to overcome thermal limitations on the beam intensity in future underground experiments. Even with best efforts in cooling, the performance of solid targets degrades under beam, which is seen for example in a reduction of target thickness or changes in the target stoichiometry. In the regime of low-energy nuclear astrophysics experiments, solid targets typically have to be replaced after an irradiation corresponding to accumulated 10$^0$-10$^1$\,(particle)~Coulombs of beam on target. This is an important practical aspect for the use of massive shielding against environmental radiation in low-background measurement. Compared to experiments on surface, where secondary cosmic radiation on shielding materials results in diminishing returns beyond a certain thickness of shielding, much more massive shielding setups of lead and copper have been used at \LUNAfourhundred{}~ (\cite{caciolli_ultra-sensitive_2009}), where for experiments with solid targets, easy access to the target had to be secured (\cite{boeltzig_improved_2018}). More sophisticated, \ie{} larger and multi-layered, shielding configurations are foreseen in the future, as a consequence of an improved understanding of the relevant backgrounds and allowed by the more spacious target station layout at the new MV facility. Target access requirements will continue to be central in future experiments with these setups that employ solid targets. 

\begin{figure}
    \centering
    \includegraphics[width=\textwidth]{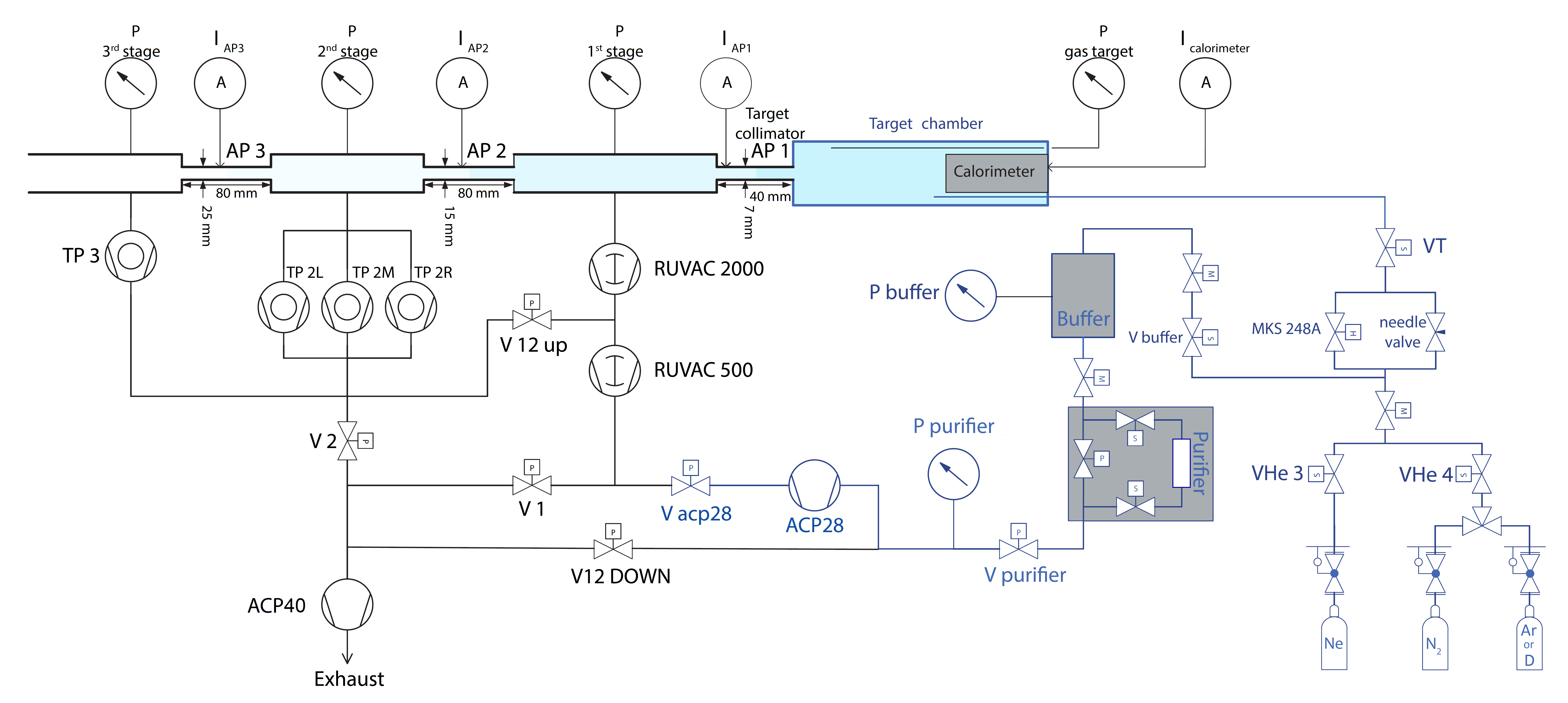}
    \caption{Differential pumping system schematic. The beam comes from the accelerator on the left, passes through the apertures AP3, AP2 and AP1, enters the target chamber and stops on the calorimeter.}
    \label{fig:GasTarget}
\end{figure}

The effects of target degradation may be avoided, wherever possible, by choosing targets in gaseous form: a windowless gas target system offers stability over the long data taking periods, up to several weeks, if needed. Another advantage is the chemical purity. Solid targets are rarely made by an element alone:  possible changes in the stoichiometry should be continuously monitored during the running time.

The gas target system presently in use at LUNA-400 accelerator is shown in Figure \ref{fig:GasTarget}. It consists of three differential pumping stages, the target chamber, the gas pipeline and a recycling system (see Figure \ref{fig:GasTarget}). 
Three pumping stages produce a strong pressure gradient between the interaction chamber and the beamline.
A water cooled collimator is placed between adjacent pumping stages, provides the correct gas flow and determines the pressure drop.
The gas target system can either recycle the gas or let it flow away.

The gas enters the interaction chamber close to the beam stop and flows into the first pumping stage, where the 99.5\% of the gas is pumped away through a roots pump.
Approximately 0.5\% of the gas also goes in the second pumping stage, where it is pumped by three turbo-molecular pumps.
A few gas flows in the third pumping stage through and is pumped away by a turbo-molecular pump.
A roots pump collects the gas from the previous pumps and is itself connected to the roughing pump or the recycling pump, depending on the running mode.

The target volume,  typically 10-40 cm long, is surrounded by the detectors and is delimited by the chamber walls, the calorimeter and the target chamber collimator.
This latter does not only collimate the beam, but also makes the pressure decrease steeply towards the first pumping stage.

The ionization of the target gas and the neutralization of the beam prevent the electrical reading of the beam current and a power compensation calorimeter with constant temperature gradient is used to monitor the beam intensity (\cite{Ferraro-2018-EPJA}).
For the proper characterization of a windowless gas target, the density and the detectors efficiency profile along the beam path must be known. The density profile is usually measured  using a mock-up scattering chamber equipped with measurement ports for capacitive pressure gauges and thermoresistors. The efficiency profile is, in turn, measured by moving radioactive sources along the beam line. Another method is the  resonance scan technique: the target system is filled with selected gases such as \nuc{14}{N} or \nuc{21}{Ne} and their narrow, strong resonances are excited with a proton beam of proper energy. The resonance position is then moved along the target by changing the beam energy accordingly. Gas target setups usually ask for heavier detector shielding systems due to the larger dimensions.

\begin{figure}
    \centering
    \includegraphics[width=0.8\textwidth]{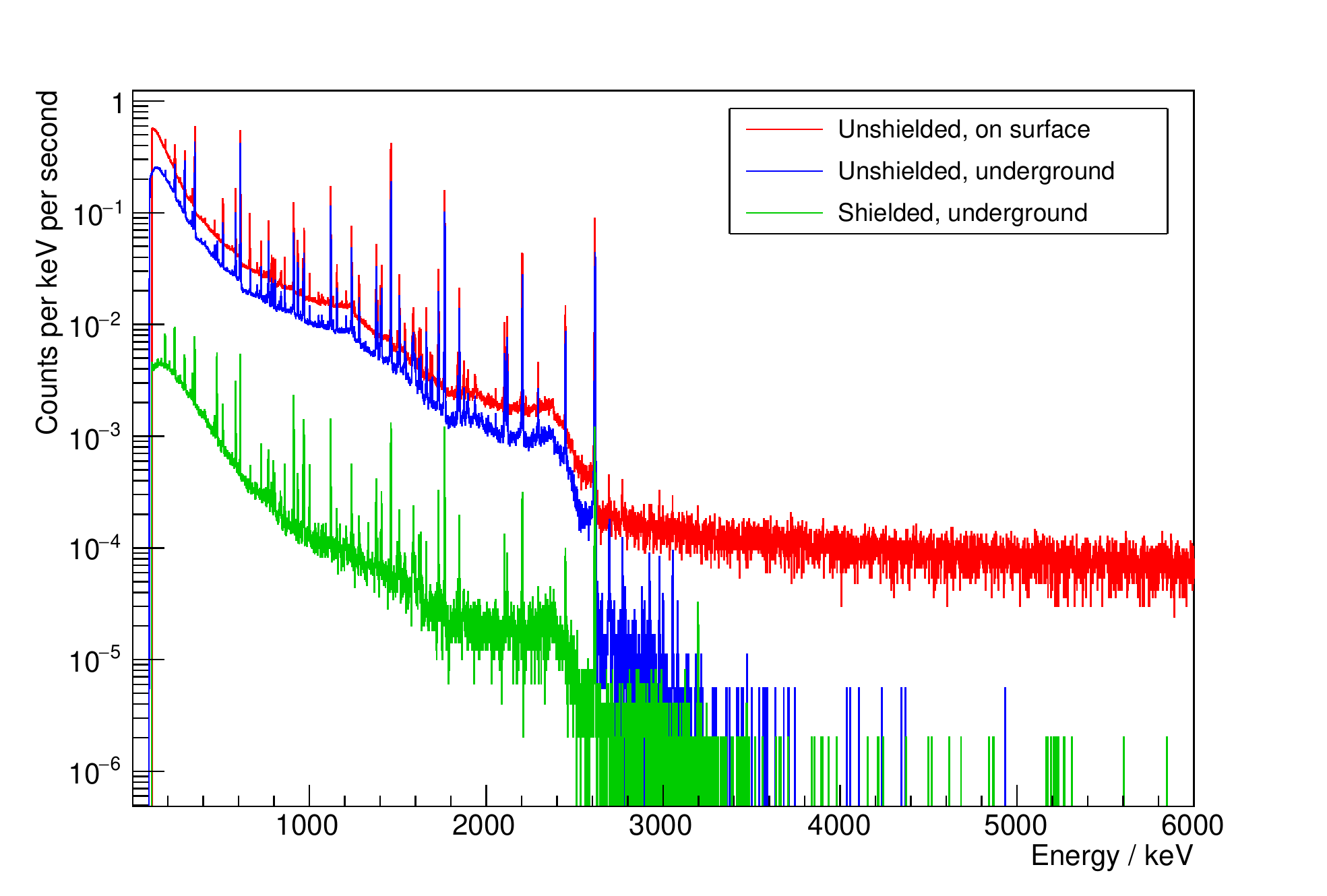}
    \caption{$\gamma$-ray background spectra taken with a Hp-Ge detector in surface laboratory (red line), at Gran Sasso (blue), and at Gran Sasso with 15 cm lead shield (green).}
    \label{fig:cac_shield}
\end{figure}

The LUNA laboratory is protected by 1400 meters of dolomite rock from cosmic ray induced effects. This rock overburden completely suppresses the hadronic and the soft electromagnetic component of cosmic rays. Muons are able to penetrate inside the mountain but their flux is  mitigated by about six orders of magnitude when compared with the Earth surface: this makes typically negligible also muon induced radiations, such as spallation neutrons or cosmogenic unstable nuclides. 
Long-lived radioisotopes such as the ones produced by the natural \nuc{238}{U} and \nuc{232}{Th} decay chains or \nuc{40}{K} are present in any laboratory and do not depend on depth, but rather on the radiopurity of rocks, buildings and detector materials. The induced gamma radiations can be mitigated by a suitable passive shielding surrounding the target and the detectors, usually consisting of selected low-background lead and freshly refined electrolytic copper. For the deep-underground setting of LUNA, a shielding of \range{15}{25}\,cm lead with low \nuc{210}{Pb} content lined at the inside with 5\,cm electrolytic copper has been found to give excellent background capabilities as shown in Figure \ref{fig:cac_shield}.
Impurities in the detector and target, on the other hand, must be minimized by proper material selection.

\begin{figure}
    \centering
    \includegraphics[scale=0.7]{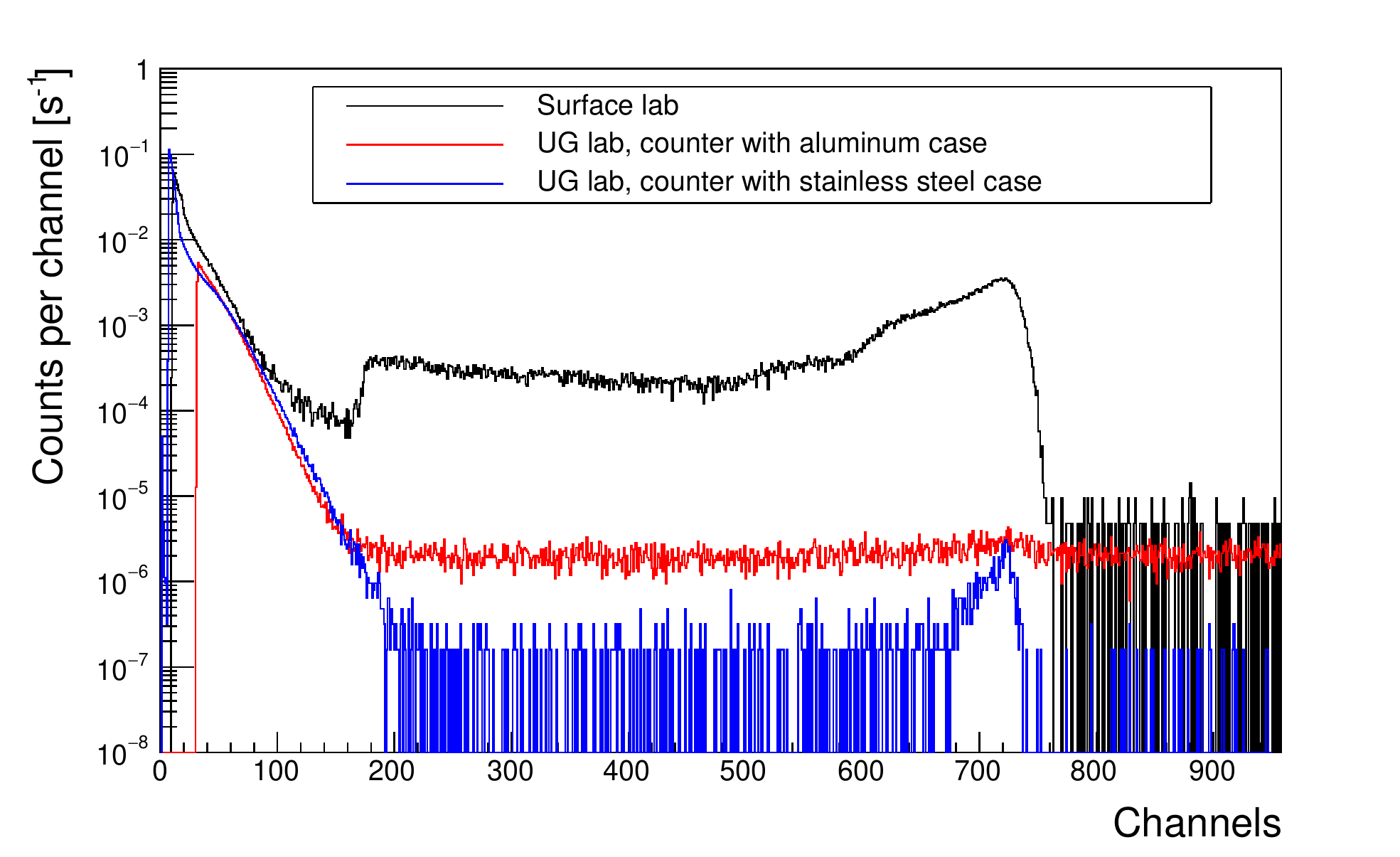}
    \caption{(Colour online) Comparison of neutron background measured by means of \He{3} counters: the black one is measured in a surface laboratory. Red and blue spectra are measured in the LNGS underground laboratory by means of counters with stainless steel and aluminum cases, respectively.}
    
    \label{fig:back_comp}
\end{figure}

 From the point of view of neutron background, the underground location allows for a reduction of 3 orders of magnitude with respect to above-ground measurements (Figure~\ref{fig:back_comp}) even without any further shielding. 
 To further increase the sensitivity in view of neutron emitting reactions that are going to be studied with the MV facility, a deep study devoted to selection material was performed to reduce intrinsic background of detectors such as \He{3} counters.
 We remind that a typical counter consists of a gas-filled tube with a high voltage applied across the anode and cathode: a neutron passing through the tube will interact with a \He{3} atom to produce tritium and a proton. These two particles ionize the surrounding gas atoms to create charges, which in turn ionize other gas atoms in an avalanche-like multiplication process.
 
Indeed, alpha particle decays, coming from impurities of uranium and thorium in the counter cases, represent the main source of intrinsic background. By selecting stainless steel cases instead of standard aluminum ones a reduction of one order of magnitude was achieved as shown in Fig.~\ref{fig:back_comp}: the blue and the red spectra were measured in the Gran Sasso with stainless steel and aluminum counters, respectively. The black spectrum is the background in a surface lab with a stainless steel counter.
As a matter of fact, the new MV facility together with the extremely low gamma and neutron background achieved by the LUNA collaboration provide a unique sensitivity to assess the key processes of post main sequence stellar burning.

\section{Neutron sources for the s-process}
The basic idea of the s-process was born in the '50s, with the famous paper by \cite{Burbidge-1957-RMP}.
It consists of a series of "slow" neutron captures and $\beta$ decays along the neutron-rich side of the valley of stability, close to the stability line.
This process is responsible for the production of about half of the elemental abundances between iron and bismuth, as stated in \cite{Kappeler-2011-RMP}, the other part being produced by the rapid neutron capture process (r-process) and to a lesser extent by the proton capture processes.

The s-process takes place in a low neutron flux, where the neutron capture rate is lower than the $\beta$ decay rate of the resulting unstable nuclei.
Such conditions are satisfied in the helium-burning shell of low-mass thermally pulsing stars in the asymptotic giant branch (main s-process) or in the helium-burning core of massive stars in the Red Giant Branch (weak s-process).
The main s-process is mostly responsible for the production of elements with $90 \le A \le 209$ (\ie{} from zirconium to bismuth), while the weak s-process contributes  to elements in the range $56 \le A \le 90$ (\ie{} from iron to zirconium).

It is well established that the \reactionCan{} reaction ($Q_\mathrm{value}=2.216\MeV{}$) is the principal neutron source for the main s-process, while the major neutron source of weak s-process is the \reactionNean{} reaction ($Q_\mathrm{value}= -0.478\keV{}$). 
The cross section of both these reactions greatly depends on temperature, the existence of excited states close to the reaction threshold and the initial abundances of the interacting species.

\subsection {The main s-process and the \texorpdfstring{\reactionCan{}}{13C(alpha,n)16O} reaction}

\cite{Kappeler-1999-PPNP} attributed the formation of the main s-process elements to thermally pulsing stars in the asymptotic giant branch (TP-AGB) with mass $1.5\,M_{\odot} < M \leq 3\,M_{\odot}$.
More recently, \cite{Cristallo-2018-APJ} indicated a slightly broader mass interval, between $1.2$ and $4\,M_{\odot}$.

\begin{figure}
    \centering
    \includegraphics[scale=0.5]{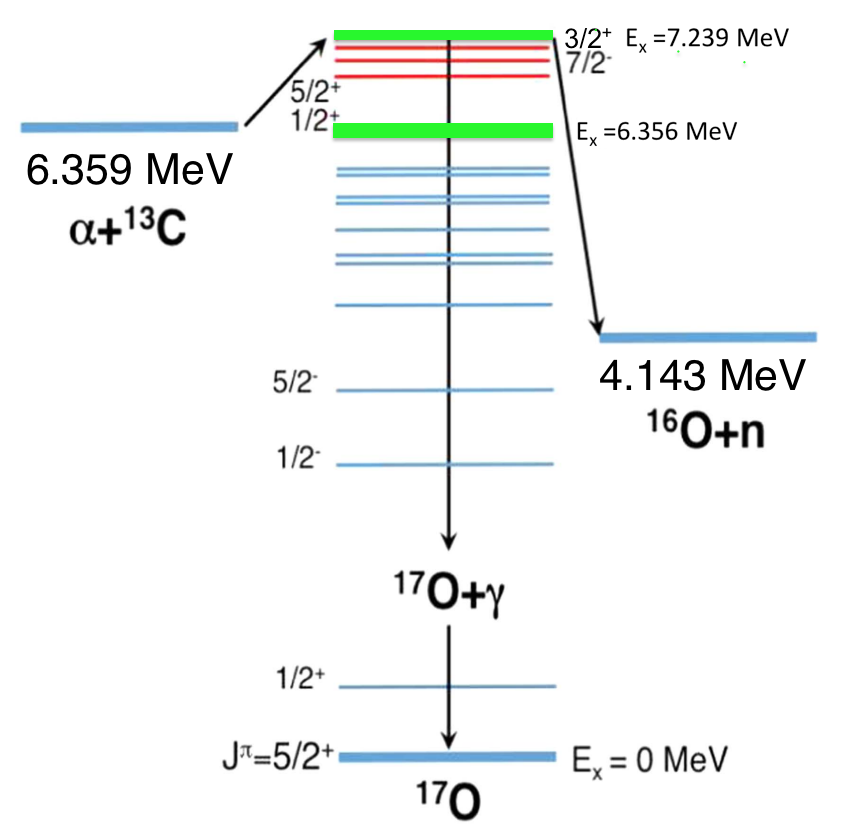}
    \caption{(Colour online) Schematic diagram adapted from \cite{Cristallo-2018-APJ} of the \reactionCan{} nuclear reaction process, together with the competing exit channel $\nuc{17}{O}+\gamma$. The excited states of interest for AGB nucleosynthesis are highlighted in green}
    \label{fig:scheme}
\end{figure}

The structure of TP-AGB stars is organized in the following layers: a carbon oxygen core, a He-burning shell, a He-rich inter-shell, a H-burning shell and a H-rich envelope.
While the H-burning shell produces helium, the core contracts and heats up the basis of the He-burning shell, whose energy production increases.
Eventually, the energy produced by the He-burning shell is not anymore radiated away efficiently and a thermonuclear runaway occurs, known as "helium shell flash" or "thermal pulse".
This translates in an expansion of the He-rich inter-shell and the cool-down of the H-burning shell, which extinguishes.
Also the He-burning shell is affected by the expansion and cools down until extinction.
A new contraction takes over and causes the initial re-ignition of the H-burning shell and of the He-burning shell afterwards, until another thermal pulse occurs.
A reservoir of \C{13}, produced via the $\C{12}(\p,\gamma)\nuc{13}{N}(\beta^+\nu)\C{13}$ reaction chain, forms the so-called \C{13} pocket at the interface between the He-rich inter-shell and the H-rich envelope.
As of today, the exact formation mechanism of such a pocket is still debated, as stated by \cite{Cristallo-2018-APJ}.
During this phase, which lasts some $10^4$ years, the \reactionCan{} reaction takes place and provides neutrons for the main s-process.\\
In the paper by \cite{Cristallo-2018-APJ}, the authors claim that, in the most metal-rich stellar models with an almost solar composition, a small amount of \C{13} might survive and be engulfed into the convective zone generated by the incoming thermal pulse.
This scenario would affect several branching points along the s-process path, and excesses of \nuc{60}{Fe}, \nuc{86}{Kr}, \nuc{87}{Rb} or \nuc{96}{Zr} are expected compared to the radiative (low neutron density) \C{13} burst.
The unburned \C{13} left at the end of the interpulse and available to produce neutrons in the subsequent pulse depends on the rate of the \reactionCan{} reaction. 
\\
The relevant astrophysical temperature for this process is $\sim$ 0.1\,GK corresponding to a Gamow energy window between 140 and 250\keV{}. Indeed, the energy range of interest could be even larger as discussed in the paper by \cite{PhysRevC.87.058801}, since the S(E) factor is energy dependent.
In Figure \ref{fig:scheme}, the level scheme of \reactionCan{} nuclear reaction process is shown. The excited states of interest for AGB nucleosynthesis are highlighted in green and red.
In particular, green levels are broad states which must be taken into account for any \reactionCan{} cross section evaluation in the astrophysical region of interest. These are the (1/2)$^+$ near threshold state and the (3/2)$^+$ at $E_{x} = 7239\keV{}$.

It is important to mention that the energy level of the near-threshold state is debated:  \cite{Ajzenberg-Selove1986} attributed to this state as a sub-threshold energy of $E_{x} = -(3 \pm 8)\keV{}$, while  recently a study by \cite{Faesterman15} deduced a positive energy value at $E_{x} = (4.7 \pm 3)\keV{}$.\\

\subsubsection {State-of-the-art}
A conspicuous number of measurements of the \reactionCan{} cross section have been carried out over the past 45 years. 

In the following we focus the attention on crucial direct and indirect measurements performed.

Among direct measurements:
\begin{itemize}
    \item \cite{Drotleff93} measured the cross section of the \reactionCan{} reaction in the \range{370}{1000}\keV{} energy range with \He{3} proportional counters embedded in a moderating polyethylene matrix. This is still the dataset with the lowest point ever measured with a direct measurement. The low-energy points reveal a S-factor enhancement, possibly due to a $1/2^+$ sub-threshold resonance mentioned by \cite{Ajzenberg-Selove1986};
    \item \cite{Brune} used \He{3} counters to measure the resonances of the \reactionCan{} reaction, at $E_\alpha = 656$ and $802\keV$: 
 the authors concluded that the resonance strengths for these two states are too weak, compared to the non-resonant contribution, to affect the stellar reaction rates;
 \item \cite{Hariss} measured  the \reactionCan{} reaction absolute cross section in an energy range $E = \range{0.8}{8}\MeV{}$ in steps of 10\keV{} with a setup similar to Drotleff's one. 
The main aim of the measurement was the geoneutrino background subtraction required by neutrino experiments such as Borexino and Kamland as explained in the paper by \cite{Araki2005}. An overall uncertainty of 4\% was achieved;
\item \cite{Heil} promoted a new study of the \reactionCan{} cross section in the energy range $E = \range{420}{900}\keV$. Heil used a different approach, employing a n-$\gamma$ converter consisting of a Cd-doped paraffin sphere surrounded with 42 $\mathrm{BaF}_2$  $\gamma$ detectors. In the central hole a neutron converter was installed. A detailed uncertainties analysis is described in the paper. The authors recognized as the main source of systematic error the change of target stoichiometry caused by the build up during the beam irradiation. At higher energies overall uncertainties could be reduced to the level of 5\%;
\item Recent measurements at high energy are due to \cite{Febbraro}, covering the same energy range spanned by Harissopulos.
They improved the precision and accuracy by means of a setup sensitive to the neutron energies, measuring also the excited state transitions via secondary $\gamma$-ray detection. With this setup, they discriminated neutrons emitted from different energy groups and they could measure the individual partial cross sections of the \reactionCan{} reaction to the ground state and second excited state of the \nuc{16}{O} final nucleus.
\end{itemize}

At low energies, uncertainties of direct measurements are larger than 50\%: they are dominated by the low counting statistics caused by unfavorable S/N ratio.\\
Moreover, going down in energy, direct measurements face limits of the fast dropping of the cross section due to the Coulomb barrier and the increase of the screening effect. 

\begin{figure}
    \centering
    \includegraphics[scale=0.75]{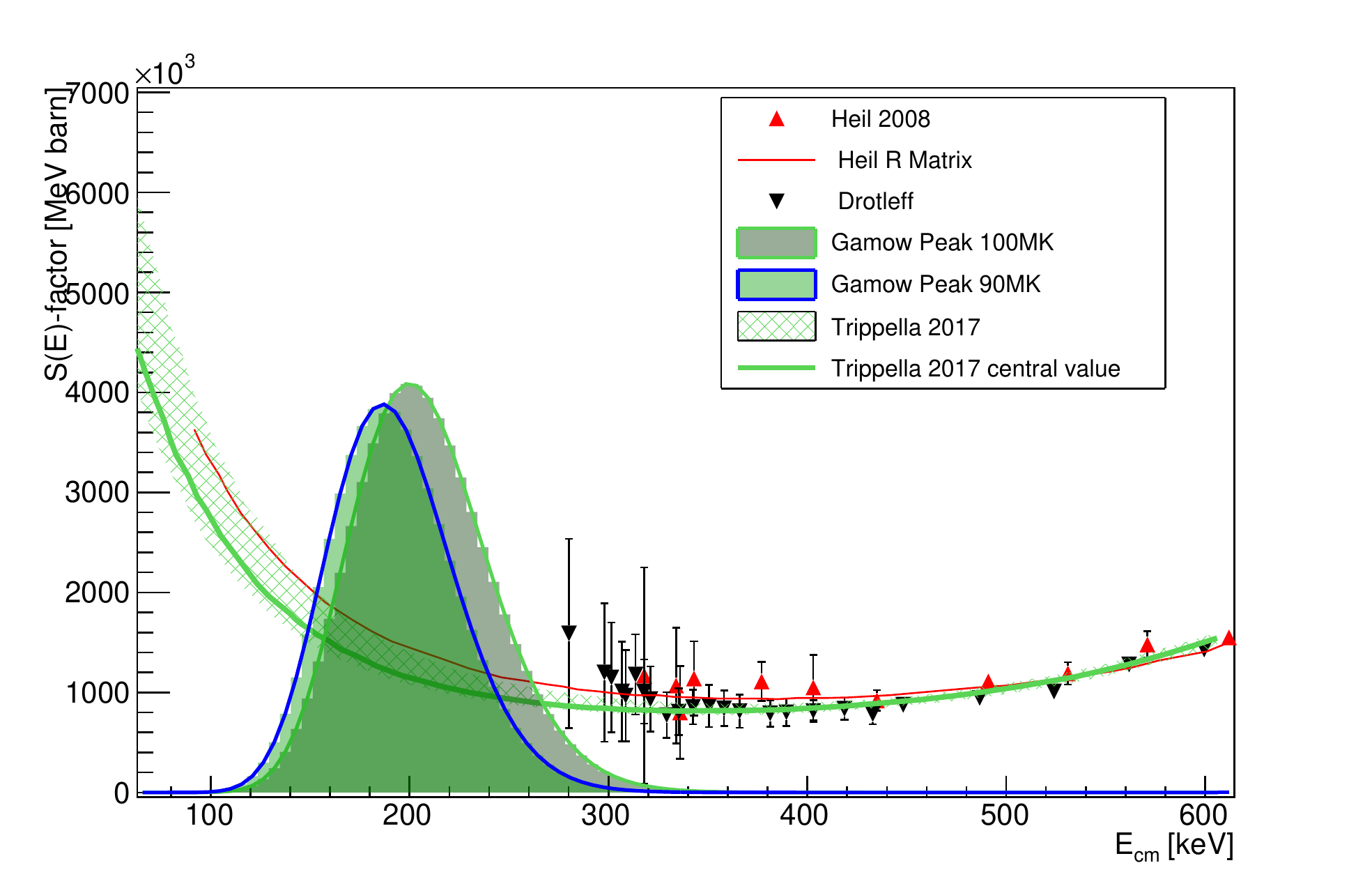}
    \caption{(Colour online) Selection of most recent \reactionCan{} measurements. Among the direct measurements the Drotleff and the Heil data are represented, indicated by black triangles and red triangles, respectively. The solid red curve indicates the R matrix extrapolation by Heil. The most recent indirect measurement by THM by Trippella et al is indicated by the green squared area and the central value is the green curve. In the plot the Gamow window for two different stellar scenarios are drawn.}
    
    \label{fig:SOA_13C}
\end{figure}

For this reason, complementary indirect studies  have been developed to better constrain the cross section of this neutron source in the relevant energy region for astrophysics. These measurements were mostly aimed to determine the spectroscopic factor and/or the asymptotic normalization coefficient (ANC) of the 1/2+ level of \nuc{17}{O} near threshold, that represents that largest source of uncertainty at low energies. 
\cite{Kubono2003} evaluated a  spectroscopic factor $S_\alpha = 0.01$, but data were reanalysed by \cite{Keeley} indicating a factor of 40 larger contribution.
The ANC method was approached for the first time in the work by \cite{PhysRevLett.97.192701} that used the $\nuc{6}{Li}(\C{13},\mathrm{d})\nuc{17}{O}$ sub Coulomb-transfer reaction. These results were recently revisited in the paper by \cite{PhysRevC.91.048801_Avila}.\\
Other indirect measurements were obtained with the Trojan Horse Method (THM): in this approach projectiles (or targets) are selected and described as clusters of two particles in quasi-free kinematics. One is involved in reaction, while the other constituent cluster, called the spectator nucleus "s",  is emitted without interacting with the system. For further information on the method one could see e.g. \cite{Tumino}. Using this technique, the $\C{13}(\nuc{6}{Li}, \mathrm{n}\,\nuc{16}{O})\mathrm{d}$ reaction was studied in quasi-free kinematic conditions (the deuteron inside the \nuc{6}{Li} beam is considered as a spectator to the three-body reaction), as described in \cite{LaCognata2013}. This work covered an energy range between $-0.3$ and $1.2\MeV{}$, and allowed to study the near-threshold resonance at $E_x = 6356\keV$. In general THM results need to be normalized to selected direct data and their uncertainty strongly depends on the choice of the reference direct measurements: in the first THM analysis by \cite{LaCognata2013}, data were scaled to the astrophysical S-factor recommended by Heil \etal{}  in the energy region between $\sim$ 0.6 and 1.2\MeV{}. As a result, a THM S-factor was good in agreement with the direct ones, but with a squared Coulomb-modified ANC $(7.7 \pm 0.3)\fm^{-1}$ not in agreement with independent assessments of the ANCs, whose weighted average is $(3.9 \pm 0.5)\fm^{-1}$.
After the new evaluation of the near-threshold resonance energy \cite{Faesterman15}, setting its center at $4.7\keV{}$ above the \C{13}–$\alpha$ threshold, data by THM where re-analyzed by \cite{Trippella} normalizing experimental data with respect to the ANC parameter of the threshold resonance obtained by \cite{PhysRevC.91.048801_Avila}. Trippella obtained an ANC value of $(3.6 \pm 0.7 \fm{}^{-1})$, in agreement with literature.\\
Data from most recent works (direct and indirect methods) are showed in Figure \ref{fig:SOA_13C}.

Both from direct measurements (high uncertainties at low energy and a large scatter in absolute values among datasets) and from indirect measurements (e.g. discrepancy in the spectroscopic factor evaluation, uncertain normalization of THM) there is a clear indication that more direct data with about 10$\%$ overall uncertainty are mandatory both at low and at high energy. 

\subsubsection{The LUNA direct measurement}
Taking advantage from the low environmental background of LNGS and the highly intense and stable alpha beam provided by the \LUNAfourhundred{} accelerator, recently the LUNA collaboration put huge efforts in the measurement of the \reactionCan{} cross section at low energy with the goal to reach an overall uncertainty near 10\%.
A detector based on 18 \He{3} counters arranged in a polyethylene moderator have been developed in order to maximise its efficiency.
\C{13} targets used during the measurement at LUNA have been produced evaporating 99\% \C{13} isotopically enriched powder on tantalum backing using the evaporator installed at the nuclear institute of research Atomki (Debrecen, Hungary). Hereby, the key points of the LUNA experiment are summarized.
As already said before, the installation of the accelerator in the LNGS underground laboratory allows a neutron background reduction of 3 orders of magnitude with respect to above-ground measurements. Moreover, a special attention was paid to reduce the $\alpha$ particle intrinsic background from detectors.\\
A further step for the background reduction was performed acquiring with Caen V1724 digitizers raw preamplifier signals from detectors and rejecting alpha signals with a pulse shape discrimination analysis described in the paper by \cite{Balibrea}. This allowed to reach an overall background in the whole detector of about 1 count/h, 2 orders of magnitude lower than previous experiments performed in surface laboratories as described in the paper by \cite{Csedreki_det}.\\
Possible beam induced background sources were investigated shooting alpha beam on blank tantalum backings. The neutron detection rate was compatible with the background measurement, making negligible the in-beam backgound.\\
The degradation monitoring under intense alpha beam is crucial during the cross section measurement performed at LUNA. The well-known NRRA (Nuclear Resonant Reaction Analysis) technique is not applicable, due to the lack of resonances in the dynamic energy range of the accelerator.
For this reason a new method of analysis was developed. \\
Data taking at LUNA consisted in long $\alpha$-beam runs with accumulated charges of $\approx 1\Coulomb$ per run, interspersed by short proton-beam runs with moderator opened and HPGe detector in close geometry, with typical accumulated charges of $0.2\Coulomb{}$ at most. During the last mentioned proton run, the target degradation can be checked perfoming a gamma shape analysis on the direct capture de-excitation to the ground state peak of \reactionCpg{} reaction with the HPGe detector.\\
Further information and details can be found in the paper by \cite{Ciani}.\\
Thanks to the unprecedented background reduction for this kind of direct measurement and the novel approach to monitor target degradation, it was possible to measure experimental yield of the \reactionCan{} reaction in an energy range between 400\keV{} down to 305\keV{} in laboratory system energy, 40\keV{} lower than data in the literature: for the first time LUNA collaboration measured with a direct technique cross section inside the Gamow window reaching unprecedented overall uncertainty ($< 20\%$).
Final results and astrophysical implication will be published within the end of 2020.\\
 The LUNA collaboration is planning to extend the  measurement of the \reactionCan{} at higher energies at the new MV facility in the LNGS laboratory. This will give the unique possibility to provide a complete data set over a wide energy range and to avoid re-normalization to other datasets with unknown systematic uncertainties.\\

\subsection {The weak s-process and the \texorpdfstring{\reactionNean{}}{22Ne(alpha,n)25Mg} reaction}

About half of the elements between iron and yttrium ($56 \lesssim A \lesssim 90$) are produced via the weak s-process in massive stars with initial mass $M > 8 M_{\odot}$ (\cite{Kappeler-RMP-2011}).
In such stars, \nuc{22}{Ne} is a by-product  of  He-burning starting from preexisting CNO isotopes.

The reaction \reactionNean{} has a negative Q-value of $-478\keV$, and requires relatively high temperatures to be ignited.
At the base of the convective envelope around the He core of massive stars, the temperature is sufficiently high  ($>$ 0.25\,GK) to make this reaction a relevant source of neutrons for the weak s-process until core He-burning extinguishes (\cite{Peters-APJ-1968,Couch-APJ-1974,Lamb-APJ-1977,Prantzos-AA-1990,Raiteri-APJ-1991b}) .
Its effectiveness as a neutron source, though,  depends also on the cross section of the competing reaction, the \reactionNeag{}.

When core He-burning runs out, \nuc{22}{Ne} is still rather abundant (about 1\% in mass as claimed in paper by \cite{Pignatari-APJ-2010}) and the reaction \reactionNean{} is reactivated during shell C-burning (\cite{Raiteri-APJ-1991a}) at a temperature of about 1\,GK.
At this stage, the \reactionCCaNe{} process yields $\alpha$ particles (\cite{Arnett-APJ-1969}) and  even larger neutron fluxes are provided as a consequence of the higher temperature.


Besides the broad interest as main neutron source in the weak s-process, it is worth mentioning some contribution also to the main s-process in low mass ($M<3M_{\odot}$) AGB stars, during thermal pulses (\cite{Gallino-APJ-1988,Hollowell-APJ-1988}), and in intermediate mass ($4M_\odot<M<8M_{\odot}$) AGB stars (\cite{Bisterzo-APJ-2014,Bisterzo-MNRAS-2015}), whereas predicted abundances of \nuc{86}{Kr}, \nuc{87}{Rb} and \nuc{96}{Zr} are at variance with observations (\cite{Lugaro-APJ-2003,GarciaHernandez-SCI-2006,GarciaHernandez-AA-2007,GarciaHernandez-APJ-2009,VanRaai-AA-2012}).

\subsubsection{State-of-the-art}
Considering the weak s-process during core He-burning, the low-energy part of the Gamow window of the \reactionNean{} reaction extends down to the boundary of the $(\alpha,\mathrm{n})$ threshold, located at $E_{\alpha,\mathrm{lab}}=575\keV$.
At such low energies, measurements have so far suffered from low signal and high background, especially because of the small cross section.
For this reason, different groups only succeeded to directly study the resonances down to $E_{\alpha,\mathrm{lab}} = 830\keV$.
Other attempts to study the resonances at lower energies by means of indirect methods, often obtained inconsistent results. In the following we summarize the most relevant direct studies of this reaction.

\begin{figure}
    \centering
    \includegraphics[width=\textwidth]{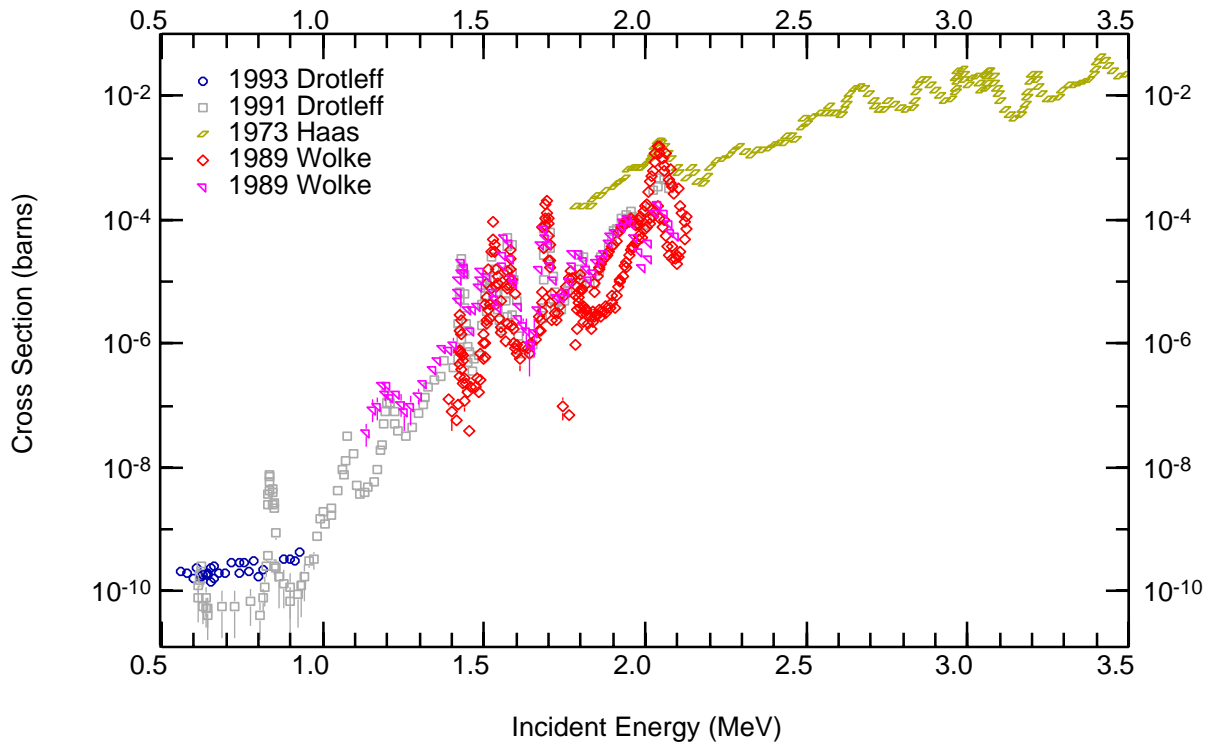}
    \caption{A subset of the previous measurements of the \reactionNean{} reaction cross section. Data retrieved from the EXFOR database version of October 8, 2020. Blue circles are upper limits.}
    \label{fig:22Ne(A,N)_XS}
\end{figure}

Back in the '60s, \cite{Ashery-NPA-1969}  discovered that it proceeds through many resonances in the compound nucleus.
Other experimental studies at about 1\MeV{} and above are due to \cite{Haas-PRC-1973}, \cite{Mak-NPA-1974} and \cite{Wolke-1989-ZPA}.

\cite{Harms-PRC-1991} investigated the energy range between $E_{\alpha,\mathrm{lab}}=0.73$ and 2.10\MeV{} with a windowless, recirculating gas target system and two \nuc{3}{He} ionization chambers in close geometry.
The resonance at $E_{\alpha,\mathrm{lab}}=830\keV$ was clearly detected but 
it was not possible to show the existence of resonances at lower energies.
Soon after, \cite{Drotleff-APJ-1993} explored a lower energy range using the same gas target and an improved 4$\pi$ detector including two concentric circles of eight \nuc{3}{He} counters in a polyethylene moderator.
Despite of the improved sensitivity, no new low energy resonances were observed in this experiment.

\cite{Giesen-NPA-1993}  performed a direct measurement with implanted \nuc{22}{Ne} targets to search for low-energy resonances.
The background from \reactionBan{}, however, limited the sensitivity at lower energies.
At the same time they investigated the excited levels with natural parity in \nuc{26}{Mg} thanks to an indirect technique, the $\alpha$-transfer.


Later, \cite{Jaeger-PRL-2001} developed a new detector with twelve \nuc{3}{He} counters arranged in an optimized geometry.
This upgrade allowed to achieve a sensitivity of $\sim$ 10\,pb and to constrain the strength of the $E_{\alpha,\mathrm{lab}}= 830\keV$ resonance to $\omega \gamma = (118 \pm 11)\mueV$. The upper limit on the tentative resonance at $E_{\alpha,\mathrm{lab}} = 633\keV$ was significantly lowered.
Based on these results, \cite{Jaeger-PRL-2001} calculated the reaction rate under the assumption that the strength of the hypothetical resonance at $E_{\alpha,\mathrm{lab}} = 633\keV$ was at 10\% of its observed upper limit. 
However, the occurrence of such a resonance
was ruled out by \cite{Longland-PRC-2009}, who demonstrated that the corresponding excited state at $E_x=11150\keV$ in \nuc{26}{Mg}
has unnatural parity.

At that time it was clear
that only  a very low-background setup in an underground laboratory could have made possible a direct investigation of the resonances at lower energies.

The focus then moved to the evaluation of the reaction rate and its implications, mostly using direct cross sections measurements at relatively high energy and indirect data.

\cite{Longland-PRC-2012} used a sophisticated statistical approach to calculate the \reactionNean{} reaction rate, including a careful treatment of the uncertainties. This led to a reduction of the uncertainties on calculated rates and raised the need for new, more precise and more sensitive measurements.

\cite{Bisterzo-MNRAS-2015} estimated the impact of the \reactionNean{} uncertainty on the isotopic abundances close to and within the branching of the s-path for main s-process.
They provided a new evaluation of the reaction rate that was a factor of 2 higher than \cite{Longland-PRC-2012}. Even if this new rate was still able to reproduce the contribution of s-only isotopes from the main s-process  within the solar uncertainties,~
\cite{Bisterzo-MNRAS-2015} underlined how a sizeable change could be caused by low-energy resonances. 

In the following years several indirect studies attempted to improve the knowledge of this reaction. A new experimental investigation by \cite{Talwar-PRC-2016} used $\alpha$ inelastic scattering to identify the important resonances and the $\alpha$ transfer technique to indirectly measure their width.
The resulting \reactionNean{} reaction rate was close to the rate in \cite{Longland-PRC-2012}.
Soon after, \cite{Massimi-PLB-2017} studied neutron capture reactions on \nuc{25}{Mg} observing several excited states of \nuc{26}{Mg} and in particular at $E_{x}=11.112\MeV$. In the same paper an R-matrix analysis was developed to assign spin and parity values to the excited states in \nuc{26}{Mg} without ambiguity and to calculate the upper limits on the reaction rates of the \reactionNean{} and \reactionNeag{} reactions. In the same work, \cite{Massimi-PLB-2017} studied the impact of these new rates on the evolution of stars with initial mass $M$ between 2 and $25 M_{\odot}$.
It was observed that for a $25 M_{\odot}$ star, the uncertainty of the \reactionNean~ reaction rate was responsible for large differences in the weak s-process abundances, up to a factor of 50 in the Sr region. Noticeable changes were also found in intermediate-mass AGB models (IMS-AGBs, $3 < M/M_{\odot} < 7$) with an increase of $\sim 50\%$ in the abundances of Y and La.

The continued interest to this reaction is demonstrated by two very recent experimental studies by \cite{Ota-PLB-2020} and \cite{Jayatissa-PLB-2020} with $\alpha$ transfer reactions: 
\cite{Ota-PLB-2020} studied the $\nuc{22}{Ne}(\nuc{6}{Li},\mathrm{d})\nuc{26}{Mg}$ in inverse kinematics, detecting outgoing deuterons and \nuc{25,26}{Mg} in coincidence.
In addition \cite{Jayatissa-PLB-2020} studied  the $\nuc{22}{Ne}(\nuc{7}{Li},\mathrm{t})\nuc{26}{Mg}$ reaction. 

\begin{figure}
    \centering
    \includegraphics[width=0.8\textwidth]{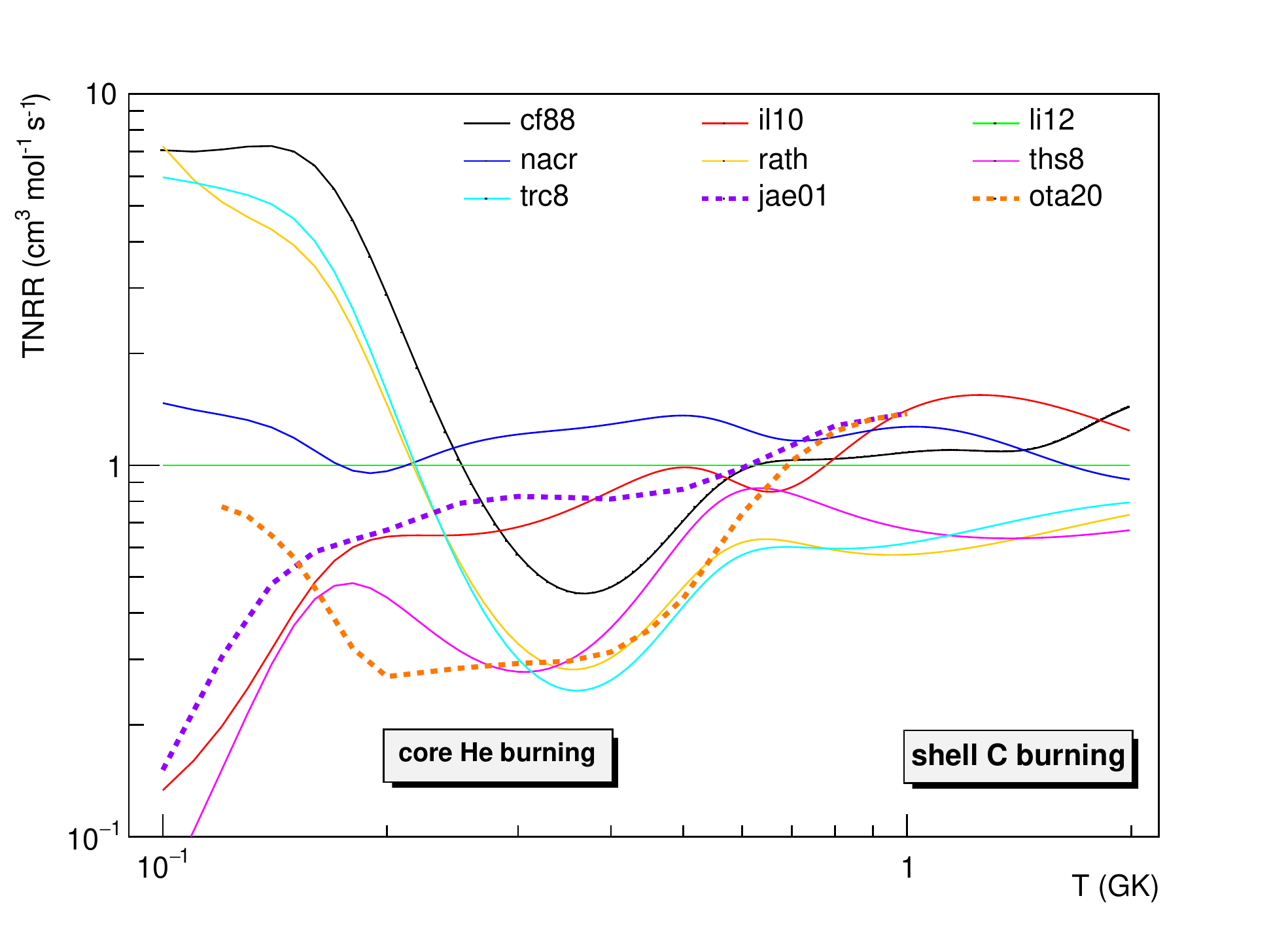}
    \caption{A subset of the \reactionNean~ reaction rate evaluations, relative to \cite{Longland-PRC-2012}. Solid lines refer to the evaluations reported in the JINA REACLIB database (\cite{Cyburt-AJSS-2010}. In particular: cf88=\cite{Caughlan-ADNDT-1988}, nacr=\cite{Angulo-NPA-1999}, rath=\cite{Rauscher-ADNDT-2000}, ths8=\cite{Cyburt-AJSS-2010}, il10=\cite{Iliadis-NPA-2010}, trc8=REFIT:\cite{Cyburt-AJSS-2010}, li12=\cite{Longland-PRC-2012}.
    Dashed lines: jae01=\cite{Jaeger-PRL-2001}, ota20=\cite{Ota-PLB-2020}.}
    \label{fig:22ne_an_25mg_relative_rates}
\end{figure}

The new evaluation of the reaction rate, based on spin-parity assignments by \cite{Jayatissa-PLB-2020} combined with data from \cite{Ota-PLB-2020}, resulted in lower rates than previous evaluations, especially at low temperatures (see Figure~\ref{fig:22ne_an_25mg_relative_rates}). The lower rate is also the result of excluding an excited state at $E_x=11.112\MeV$, corresponding to $E_{\alpha,\mathrm{lab}}=598\keV$, observed by \cite{Massimi-PLB-2017} and not observed in these studies.

In conclusion the thermonuclear reaction rate of this reaction is still largely uncertain: several evaluations are present in the literature (see Figure \ref{fig:22ne_an_25mg_relative_rates}), based on theoretical considerations, direct and indirect measurements, differing up to a factor of 5 in the temperature range relevant to the s-process in core He burning.
The presence of low-energy resonances in  the \reactionNean~ reaction below $E_{\alpha,\mathrm{lab}} = 830\keV$ are expected, based on known levels in \nuc{26}{Mg}, but no such resonances have been directly observed, yet. Nevertheless they might contribute significantly to the reaction rate around 0.2\,GK and cause sizeable changes in the prediction of weak s-process abundances.

The direct measurement of the \reactionNean~ reaction cross section will be carried out at the new MV facility at LNGS (\cite{Guglielmetti-EPJWC-2014,Prati-NICXV-2019}), using a windowless gas target (see Fig.~\ref{fig:GasTarget}) of enriched \nuc{22}{Ne}.
Such experiment could provide precise and accurate cross section measurements down to about $E_{\alpha,\mathrm{lab}} \sim 600\keV$.
Most of the background is expected to be due to the \reactionBan{} reaction, as already reported by past experiments and therefore a proper reduction of contaminants poses a crucial challenge, combined with the development of an optimized detector setup.

SHADES (Scintillator-He3 Array for Deep underground Experiments on the S-process) is an ERC starting grant (Grant agreement ID: 852016), recently awarded to realize a new setup for the measurement of the \reactionNean~ reaction at energies of astrophysical interest.
SHADES includes the development of a novel neutron detector and a gas target to be used at LUNA.
The detector combines an array of \nuc{3}{He} counters with their high detection efficiency and liquid scintillators, which act as moderators for the reaction neutrons while at the same time providing information on the neutron energy.
The combination of \nuc{3}{He} tubes and scintillator, together with recently studied signal processing techniques, see \cite{Balibrea}, will be able to limit backgrounds from external and internal sources as well as beam-induced background to acceptable levels.
The new detector will allow an increase of at least two orders of magnitude in sensitivity, allowing for the first time a measurement of the reaction cross section in the energy range relevant to the s-process in core He burning.

\section{The \texorpdfstring{\reactionCag{}}{12C(alpha,gamma)16O} reaction}

The reaction \reactionCag{} competes with the so-called triple-$\alpha$ process ($\He{4} + \He{4} \to \nuc{8}{Be}$ followed by $\He{4} + \nuc{8}{Be} \to \C{12}$) during stellar helium burning (\cite{Burbidge-1957-RMP}). The astrophysical rates of both reactions influence the ratio of \C{12}/\nuc{16}{O} produced during the helium burning phase, which in turn determines following steps of stellar evolution. Due to the central role of these nuclides, understanding their ratio in helium burning has been identified as a problem of ``paramount importance" \cite{fowler_quest_1984} for nuclear astrophysics. Compared to the triple-$\alpha$ process, the cross section of \reactionCag{} is significantly less well-known and, in spite of extensive experimental efforts, a better understanding of this reaction remains desirable. A recent comprehensive review on the state of understanding \reactionCag{} can be found in \cite{deboer_12c16o_2017}.

Owing to the sharp drop of the charged particle reaction cross sections towards the energy regions relevant for astrophysics, direct measurements in the energy region of interest are not available, making extrapolations necessary. Such extrapolations are challenging due to the nuclear structure of the compound nucleus \nuc{16}{O}: the cross section in the energy range of interest is characterized by the presence of broad resonances (including sub-threshold states). It is crucial to study the interference between states of the same J$^{\pi}$, but also to account for angular effects from the interference of processes with different $J^\pi$ (as outlined in \cite{deboer_12c16o_2017}).  In particular, the E1 and E2 components of capture to the ground state are of comparable strength in the energy range of interest, and the extrapolated cross section is very sensitive to the interference of these two components.

Different experimental approaches have been taken to directly study the \reactionCag{} reaction: in normal kinematics, a fixed \C{12} target (solid or gaseous) is bombarded by $\alpha$ particles, detecting $\gamma$-rays from the reaction; inverse kinematics employs a \C{12} beam impinging on a helium target. Inverse kinematics experiments have been performed as measurements of the $\gamma$-rays from the reaction, or detecting the \nuc{16}{O} nuclei in a recoil separator (\cite{kremer_coincidence_1988,schurmann_first_2005,matei_measurement_2006,schurmann_study_2011}). Studies of the inverse reaction $\nuc{16}{O}(\gamma_0,\p)\C{12}$ at high-intensity $\gamma$-ray facilities 
allow to infer information on the ground state transitions. Other reactions to study the nuclear structure of \nuc{16}{O} are used to constrain extrapolations of the reaction \reactionCag{} in frameworks such as \emph{R}-matrix theory. \cite{deboer_12c16o_2017}.

\begin{figure}[htb]
	\centering
	\includegraphics[width=0.8\textwidth]{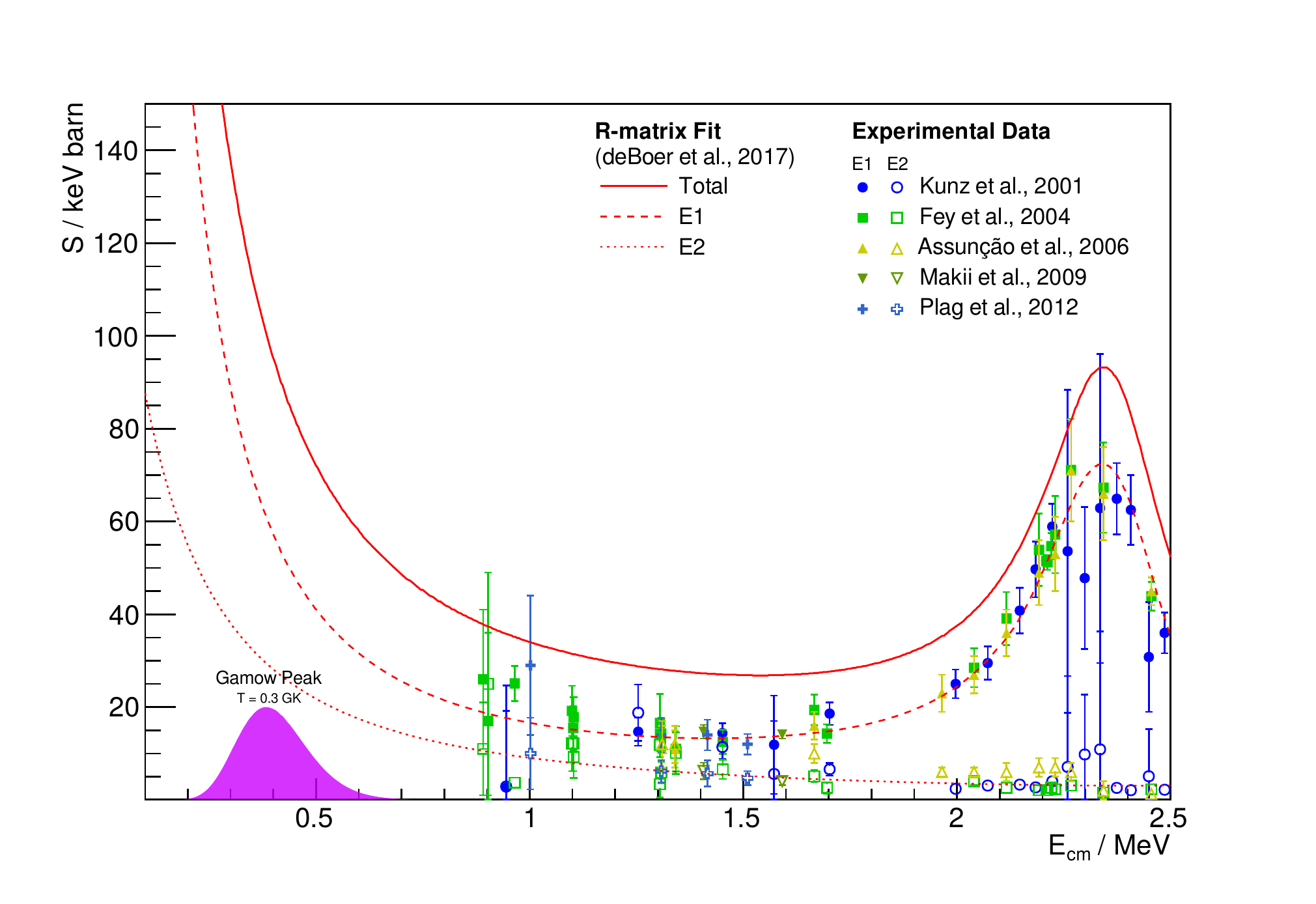}
	\caption{Overview of recent experimental data \textcolor{red}{(}\cite{kunz_12c16o_2001,fey_im_2004, assuncao_e1_2006, makii_e1_2009,plag_12c16o_2012}\textcolor{red}{)} for the ground state capture in \reactionCag{}, with the results of R-matrix fits from \cite{deboer_12c16o_2017} for comparison. All data is unscaled. The location of the Gamow window for a stellar temperature of 0.3\,GK is shown for reference.}
	\label{fig:12Cag}
\end{figure}

When measuring the $\gamma$-rays from reaction, angular distribution measurements at multiple detector angles yield information to disentangle the E1 and E2 components, whilst a total absorption spectroscopy setup, detecting the total $\gamma$-ray energy, yields the total cross section with a large detection efficiency. Figure \ref{fig:12Cag} summarizes the current situation of direct measurements, showing recent direct measurements of this reaction at low energies for illustration. 
These measurements extend down to about 0.9\,MeV center of mass energy, but are characterized by increasing uncertainties when approaching these low energies. As the cross section drops rapidly towards these energies, backgrounds -- environmental and beam-induced -- are increasingly relevant. For example experiments in normal kinematics are affected by backgrounds from the reaction \reactionCan{}, which has a cross section that is of the order of $10^6$ times that of \reactionCag{}. Neutrons can produce background signals directly in the detector, or through secondary radiation in the environment of the detector. This background can be reduced by using \C{12} targets depleted in \C{13}, or with the help of bunched beams that allow to disentangle the prompt $\gamma$-ray signal from neutron-induced backgrounds by time-of-flight.

Additional data at lower energies is desirable to better constrain the energy-dependence of the extrapolation, and further experiments will aim to shed light on it in the future. Direct measurements are expected to contribute to this effort by pushing the lower limit for the available cross section data further below 1\,MeV center of mass energy. This includes promising measurements with a recoil mass separator \cite{fujita_direct_2015}. On the side of the new underground accelerator facilities, new exciting opportunities for the study of this reaction will become available shortly. Measurements of \reactionCag{} are among the scientific goals of the new MV facility at LNGS and the Felsenkeller shallow-underground accelerator laboratory for nuclear astrophysics, \cite{bemmerer_felsenkeller_2018}. Both accelerators will not only provide beams of $\alpha$ particles, but also of carbon ions, allowing for underground measurements of this reaction in inverse kinematics. The scientific program of JUNA at JPL, as outlined in \cite{liu_underground_2017}, includes the study of \reactionCag{} as well. To take full advantage of the high-intensity $\alpha$ beam and the deep underground location of JUNA, the minimization of beam-induced backgrounds, such as those created on \C{13}, has been identified as crucial.

\section{ The \texorpdfstring{\reactionCC{}}{12C+12} reaction}


At the end of the core helium burning, the central part of the star becomes more massive, contracts and heats up. The contraction and the possible consequent collapse can be halted
by the ignition of carbon burning or by the pressure of degenerate electrons. There are several factors preventing the ignition temperature for carbon burning is reached prior to electron degeneracy. For instance, plasma neutrinos are produced near the center of the star and they cause a decrease of the central temperature while leaving it. In addition, in the case of intermediate mass stars, the second dredge-up further reduces the temperature of the star core, with the convective envelope penetrating into the H-exhausted shell. Depending on the star mass, it may attain the physical condition for C burning or become a carbon-oxygen white dwarf. The minimum initial mass of a star able to experience a C-burning phase is called $M_{\rm up}$. The value of $M_{\rm up}$ was proposed for the first time by \cite{Becker-80-ApJ} who found $M_{\rm up} = 9 M_\odot$ for a star with nearly solar composition. However, there are many uncertainties: those affecting the \reactionCC{} and $\C{12}+\alpha$ rates are the most important nuclear ones. As a matter of fact, the value of $M_{\rm up}$ separates the progenitors of C-O white dwarfs, novae and type Ia supernovae, from those of core-collapse supernovae, neutron stars and stellar mass black holes. Finally, if the star mass is slightly higher than $M_{\rm up}$, an off-center carbon ignition takes place in degenerate conditions and the star may end its life as a O-Ne white dwarf.

Stellar models predict that carbon burning, triggered by the \reactionCC{}, occurs for center of mass energies between 0.9 and 3.4\MeV{}. The reaction can proceed through different channels corresponding to the emission of a photon, a neutron, a proton, one or two $\alpha$ particles or a \nuc{8}{Be} nucleus. Among these channels, the two \textcolor{red}{most} relevant are the \reactionCCpNa{} and \reactionCCaNe{}; alpha particles can produce neutrons through \reactionCan{} and \reactionNean{} reactions. These neutrons are fundamental for the synthesis of elements heavier than Fe through the s-process. 

The \reactionCC{} reaction rate at center of mass energies $\approx 1.5\MeV{}$ also affects the physical conditions in the SNIA explosion. In particular, carbon burning can be ignited in explosive condition when material is accreted on the surface of a white dwarf in a close binary system \cite{Bravo-11}. A variation in the rate would modify the extension of the convective core prior to the explosion, the degree of neutronization and the temperature at the beginning of the thermonuclear runaway. The knowledge of SNIA is fundamental in cosmology since these systems allow the measurements of distances and of the expansion rate of high redshift galaxies \cite{Tutusaus-19}.

Unfortunately the Gamow window of the \reactionCC{} reaction, $\range{0.7}{3.4}\MeV{}$ depending on the astrophysical scenario, is much lower than the height of the Coulomb barrier, $6.7\MeV{}$ approximately, making the direct measurement of the cross section extremely difficult. 

\subsection{State-of-the-art}

The two most relevant channels in the \reactionCC{} reaction are the emission of protons and $\alpha$ particles, with a Q-value of $2.24\MeV{}$ and $4.62\MeV{}$, respectively. The proton and alpha channels can be measured either by detecting the charged particles or the gamma decay. In particular, the largest branching is for the dexcitation of the first excited state to the ground state of the \nuc{23}{Na} or \nuc{20}{Ne}. Above 2 MeV, the first excited state transition to the ground state accounts for approximately 50\% of the total cross section and produces photons of 440\keV{} and 1634\keV{} in the case of proton or alpha emission, respectively.

The challenge in obtaining a reliable measurement of the \reactionCC{} cross section at low energies, is related to its exponentially falling behaviour which produces a very low counting rate; in this scenario any natural or beam-induced background can seriously affect the measurement. The latter is due to impurities in the carbon target, manly hydrogen and deuterium, because they can form bonds with carbon. The main background related to the gamma measurements comes from the $\nuc{2}{H}(\C{12},\mathrm{p}_1\gamma)\C{13}$ and $\nuc{1}{H}(\C{12},\gamma)\nuc{13}{N}$ reactions, as detailed in the experimental work by Spillane \etal{}. The Compton background of the primary peaks could completely dominate the carbon fusion $\gamma$-ray peaks \cite{Spillane-08-PRL}. As far as the particle measurements are concerned, it is kinematically impossible to find protons in the carbon fusion region of interest, if the particle detectors are placed at backward angles.

In the following the most recent papers focused on the \reactionCC{} cross section measurement at low energies are summarized.

\cite{Jiang-18-PRC} have recently measured the \reactionCC{} fusion cross section in the energy range $\range{2.5}{5}\MeV{}$. The authors studied the two main channels: \reactionCCpNa{} and \reactionCCaNe{} at Argonne National Laboratory using a Gammasphere array of 100 Compton-suppressed Ge spectrometers in coincidence with silicon detectors. The measurement was pushed down to $2.84\MeV{}$ and $2.96\MeV{}$ for the p and $\alpha$ channels, respectively; the results are in good agreement with other measurements using $\gamma$ \cite{Spillane-08-PRL} and charged particle detection \cite{Zickefoose-18-PRC}, but with smaller uncertainties.

\cite{Tumino-18-Nature} measured the cross section of the \reactionCCpNa{} and \reactionCCaNe{} reactions through the indirect Trojan Horse Method (THM). A $30\MeV{}$ beam was delivered on a natural carbon target; charged particles were detected through $\Delta$E-E position sensitive silicon detectors. The THM results for $\alpha$ and p channels are in good agreement with direct data except for the $2.14\MeV{}$ region, where the claim of a strong resonance by previous works \cite{Spillane-08-PRL} is not confirmed. Instead the indirect data show a resonance at $2.095\MeV{}$, one order of magnitude less intense with respect to the $2.14\MeV{}$ resonance found by Spillane in the $\nuc{20}{Ne} + \alpha$ channel and of similar intensity in the $\nuc{23}{Na} + \p{}$ one. In addition, several low-energy resonances are evident below $1.5\MeV{}$, never detected before in a direct measurement. The results of the THM raised some criticism \cite{Mukhamedzhanov-19-PRC} mainly because of the neglected Coulomb interaction between \nuc{2}{H}, the spectator nucleus in the THM, and $\nuc{24}{Mg}$.

The \reactionCCpNa{} has also been measured by \cite{Zickefoose-18-PRC} in the $\range{2}{4}\MeV{}$ energy range by particle spectroscopy. The beam, provided by the tandem accelerator of the Center for Isotopic Research on the Cultural and Environmental (CIRCE) heritage, was sent onto highly ordered pyrolytic graphite targets; protons were detected through $\Delta$E-E silicon detectors. The total S-factor, including also the contribution of the $\alpha$ channel, has been obtained using the ratio between the p-channel and total S-factor provided by \cite{Becker1981}. Due to the poor statistics and beam induced background problems, this work needs a further experimental effort to improve the knowledge of the total S-factor in the relevant energy range. For this reason the experimental campaign continued with a new study devoted to the reduction of light species contaminant, especially \nuc{1}{H} and \nuc{2}{H} in the carbon targets \cite{Morales-18-EPJA}. Measurements were done with natural graphite and highly ordered pyrolytic graphite targets. \nuc{1}{H} and \nuc{2}{H} content were reduced up to 70-85\% by means of diffusion at high temperatures (higher than $1000\degC{}$). A further reduction of a factor of 2.5 was obtained enclosing the scattering chamber in dry nitrogen to minimize leaks into the rest gas within the chamber. The bulk contamination finally achieved by the authors is 0.3\,ppm. Further measurements are planned with the new experimental setup.

An upper limit on the \reactionCC{} S-factor has been recently suggested from the measurement of the $\C{12}+\C{13}$ reaction by \cite{Zhang-20-PLB}; in fact it has been observed that the $\C{12}+\C{13}$ and $\C{13}+\C{13}$ cross sections at energies below and above the Coulomb  barrier are upper bounds of the non-resonant contribution of the \reactionCC{} cross section. The measurement of the  $\C{13}+\C{13}$ reaction was performed by studying the  $\C{12}(\C{13},\p{})\nuc{24}{Na}$ channel; \nuc{24}{Na} has an half life of 15.0~hours, allowing an activation measurement. The resulting upper limit on the \reactionCC{} S-factor agrees nicely with available direct experimental data down to $\approx 2.5\MeV{}$, while for lower energies the THM results are significantly higher compared to the Zhang upper limit. However, this result should be taken with caution considering that the obtained upper limit is only valid for the non-resonant component of the \reactionCC{} cross section.
Recent theoretical calculations of the \reactionCC{} sub-barrier fusion cross section highlighting the role of resonances can be found in \cite{bonasera2020calculation}.

Another step forward in the knowledge of the \reactionCC{} rate has been recently moved by \cite{Fruet-20-PRL}. They performed a direct measurement down to $\approx 2.2\MeV{}$ using the particle-gamma coincidence technique. The experiment was performed at the Andromede accelerator facility at IPN Orsay, France with a \C{12} beam, maximum beam current of 2\,p$\mu$A for astrophysically relevant energies, impinging on a natural carbon target. Charged particles were detected through three annular silicon strip detectors covering 30\% of the total solid angle. For gamma-ray detection, an array of LaBr$_3$(Ce) scintillator detectors has been employed. The results are in good agreement with the data reported by \cite{Jiang-18-PRC} and with \cite{Tumino-18-Nature}. However, a more prominent resonance has been observed around $3.8\MeV{}$ compared to other measurements ( \cite{Spillane-08-PRL,Zickefoose-18-PRC}).

The most recent measurement of the \reactionCC{} cross section has been performed by \cite{Tan-20-PRL} at the University of Notre Dame. The simultaneous detection of protons and alphas, through a silicon detector array, and $\gamma$-rays with a 109\% HPGe detector, allowed for particle-$\gamma$ coincidence technique. The S-factor upper limit at $2.2\MeV{}$ for proton (p$_1$) and alpha ($\alpha_1$) channels are lower than THM data. We note that the upper limit for the proton channel disagrees significantly with the recent measurement of \cite{Fruet-20-PRL}. The discrepancy is less evident, but still present, for the alpha channel. In the energy region between $2.5$ and $3\MeV{}$, there is some tension between the results of \cite{Tan-20-PRL} and previous measurements \cite{Jiang-18-PRC} both for proton and alpha channel. The S-factor results at center of mass energies above $4\MeV{}$ agree nicely with other data.

\begin{figure}
    \centering
    \includegraphics[width=0.9\textwidth]{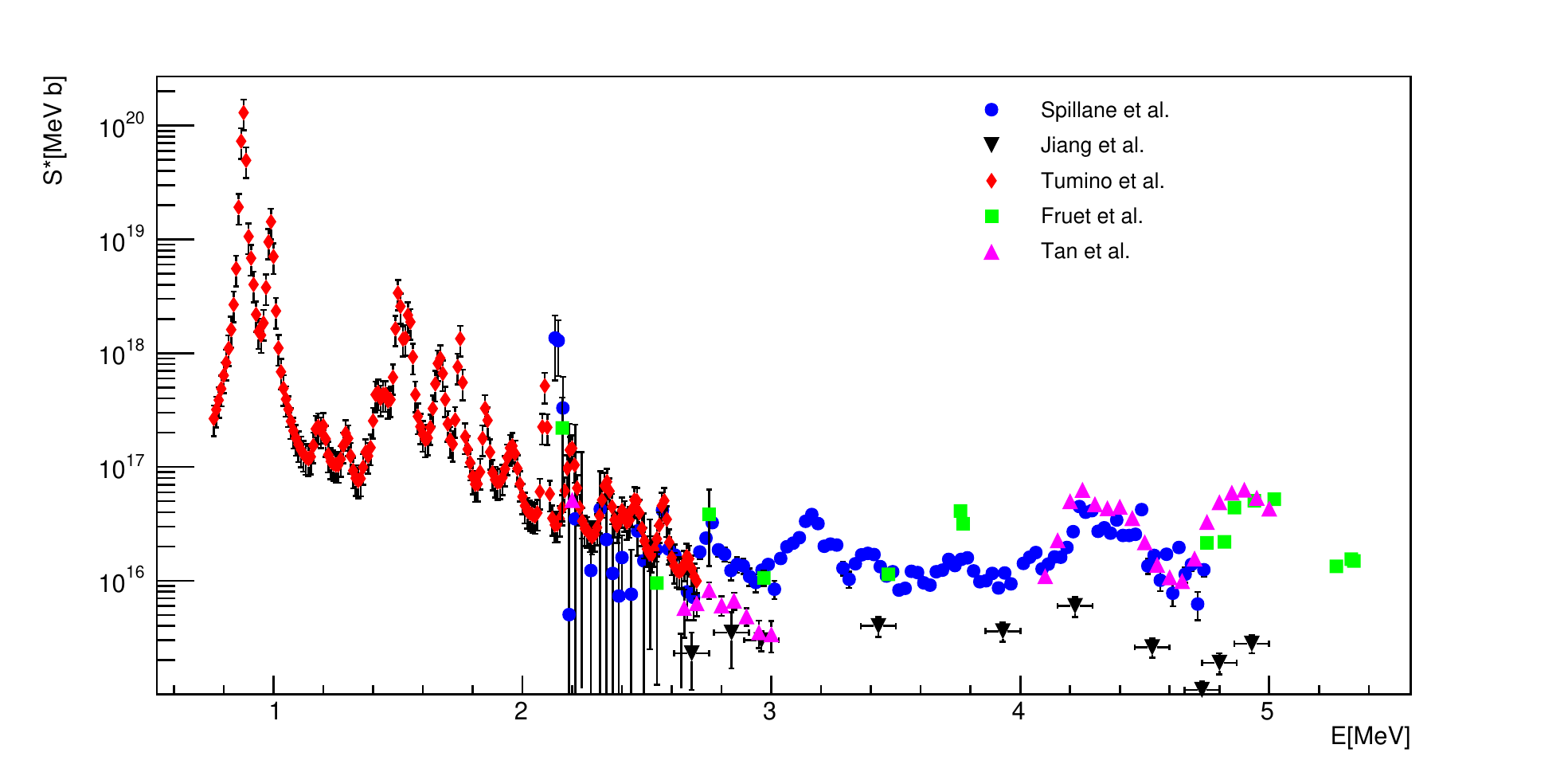}
    \caption{S-factor values obtained by \cite{Spillane-08-PRL}, \cite{Jiang-18-PRC}, \cite{Tumino-18-Nature}, \cite{Fruet-20-PRL} and \cite{Tan-20-PRL}}
    \label{fig:12c_summary}
\end{figure}

A comparison between the total S-factor values obtained by \cite{Spillane-08-PRL}, \cite{Jiang-18-PRC}, \cite{Tumino-18-Nature}, \cite{Fruet-20-PRL} and \cite{Tan-20-PRL} is shown in Figure \ref{fig:12c_summary}. It should be underlined that \cite{Tumino-18-Nature} data are normalized to direct measurements, so a difference in the absolute value of the S factor can also be attributed to systematic errors affecting direct data. Significant discrepancies between the results of the reported experiments are evident in whole energy range and, for this reason, a further experimental effort is needed.

\subsection{The measurement in an underground laboratory}
\begin{figure}
    \centering
    \includegraphics[width=0.8\textwidth]{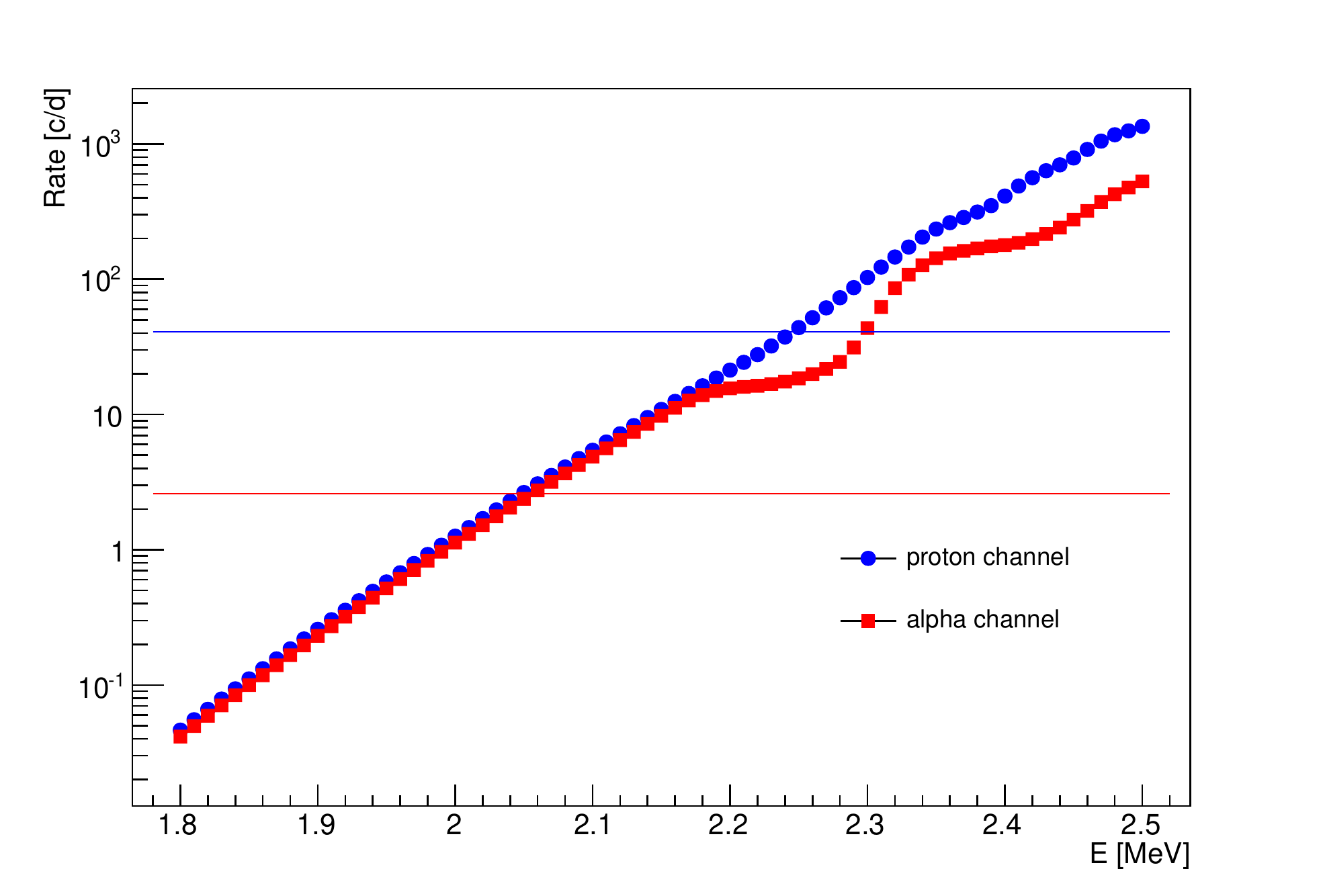}
    \caption{Counting rate, in counts per day, obtained considering data provided by \cite{Spillane-08-PRL}, a $\gamma$ detection efficiency of 6\% and 2\% for 440\keV{} and 1636\keV{} $\gamma$ energies, respectively and a beam current of 50-150 $\mu$A. The two horizontal lines represent a typical rate of $\gamma$ background measured at LNGS with a shielded setup \cite{caciolli_ultra-sensitive_2009} (blue and red lines for 440\keV{} and 1636\keV{} $\gamma$ energies, respectively)}
    \label{fig:12c_rate}
\end{figure}

An underground location such as the one of the LUNA experiment is the perfect environment to perform the measurement of the \reactionCC{} cross section detecting $\gamma$-rays emitted in the decay of the \nuc{23}{Na} and \nuc{20}{Ne} excited states. A high-efficiency and ultra-low intrinsic background germanium detector (HPGe) is suitable for the measurement in combination with a massive lead shielding to avoid the contribution of the low-energy gamma-rays coming from the decay of the \nuc{238}{U} and \nuc{232}{Th} chains. In figure \ref{fig:12c_rate} the counting rate, expressed in counts per day, is reported as a function of the interaction energy. To calculate the rate, the S-factor provided by \cite{Spillane-08-PRL} has been adopted, considering that the decay of the first excited state to the ground state accounts for $\approx$ 50\% of the total cross section and produces photons of 440\keV{} and 1634\keV{} in the case of proton or alpha emission, respectively. It is evident that if the trend of the S-factor observed by \cite{Tumino-18-Nature} is confirmed the reaction rates can be higher by 1-3 order of magnitude. The two horizontal lines represent a typical rate of $\gamma$ background measured at LNGS with a shielded setup \cite{caciolli_ultra-sensitive_2009} (blue and red line for 440\keV{} and 1636\keV{} $\gamma$ energies, respectively). \textcolor{red}{In particular for the proton channel, crucial issues are the choice of the materials to limit the intrinsic contaminants  and a proper detectors shielding. In addition, constant nitrogen fluxing around the setup could help to further reduce the background, avoiding radon contaminants. The $\gamma$-detection efficiency adopted in the calculation is just a standard value, new high efficiency setups will be developed for the future measurements.
From a rough estimation considering the data provided by \cite{Spillane-08-PRL} and the setup described in Figure \ref{fig:12c_rate}, we can say that the dominant contribution to the background for the proton channel will come from the environmental radioactivity if  a 0.3 ppm H contamination level is achieved in the targets (\cite{Morales-18-EPJA}) making the induced background not an issue at least down to $\sim$ 2 MeV. The limitation in the alpha channel, is conversely related to the low rate. }
To provide the total cross section, the measurement of the charged particle channels is also needed. In this case the advantage of the underground location is less evident but still present; in fact  secondary particles produced by the passage of cosmic rays through the detectors could contribute to the background and they are effectively reduced at LNGS (\cite{Bruno2015}).

\section{Conclusions}
The enhancement in the sensitivity provided by the strong background reduction in a underground laboratory, together with the best experimental techniques, have made possible, during twenty-five years of LUNA activity, to take clear steps forward in the knowledge of nuclear processes relevant to astrophysical scenarios.
The installation of a new MV accelerator in the Gran Sasso laboratory will allow over a broad time window of at least twenty years, to extend these studies to key processes of  helium, carbon and neon burning phases. 
Even if more extensively studied, also other important processes of H-burning will be better constrained thanks to the new facility. An example is the \reactionNpg{} reaction, presently known only at energies well above the Gamow peak. By combining the existing LUNA 400\,kV machine with new LUNA-MV facility it will be possible  to cover the necessary energy range with a sufficient overlap and without any hole between 200\keV{} and 1.5\MeV{}, allowing to reduce the systematics in the extrapolations down to the 5\% level. The \reactionNpg{} reaction will also be suitable to perform the commissioning and the tuning of LUNA MV accelerator.

As already said, the success of the LUNA approach has motivated similar facilities already in operation in the United States or under construction in the Republic of China. This worldwide effort will allow in the next decades to take important steps forward in the field of nuclear astrophysics.

\bibliographystyle{frontiersinHLTH&FPHY} 
\bibliography{references}

\end{document}